\documentclass[a4paper, 10pt]{article}
\usepackage{bm}
\usepackage{geometry}
\usepackage{abstract}
\usepackage{lmodern}
\usepackage{amsmath}
\usepackage{xcolor}
\usepackage[table]{xcolor}
\usepackage{graphicx}
\usepackage{booktabs}
\usepackage{authblk}
\usepackage{nameref}
\usepackage{xurl}
\usepackage[colorlinks=true, allcolors=blue]{hyperref}
\usepackage[english]{babel}
\usepackage[utf8]{inputenc}
\usepackage{multirow}
\usepackage{float}
\DeclareUnicodeCharacter{0301}{\textsuperscript{\char"00B4}}
\DeclareUnicodeCharacter{0300}{}
\renewcommand{\arraystretch}{1.0}

\begin{document}

\title{Going beyond density functional theory accuracy: Leveraging experimental data to refine pre-trained machine learning interatomic potentials}

\author[1,2]{Shriya Gumber}
\author[1]{Lorena Alzate-Vargas}
\author[1]{Benjamin T. Nebgen}
\author[3]{Arjen van Veelen}
\author[4]{Smit Kadvani}
\author[1]{Tammie Gibson}
\author[*1,5]{Richard A. Messerly}

\affil[1]{Theoretical Division, Los Alamos National Laboratory, Los Alamos, NM 87545, USA}
\affil[2]{Department of Chemistry, University of Southern California, Los Angeles, CA 90089, USA}
\affil[3]{Material Science and Technology Division, Los Alamos National Laboratory, Los Alamos, NM 87545, USA}
\affil[4]{Department of Quantitative and Computational Biology, University of Southern California, Los Angeles, CA 90089, USA}
\affil[5]{National Center for Computational Sciences Division, Oak Ridge National Laboratory, Oak Ridge, TN 37830, USA}
\affil[*]{Corresponding author: messerlyra@ornl.gov}

\date{}
\maketitle

\section*{Abstract}

Machine learning interatomic potentials (MLIPs) are inherently limited by the accuracy of the training data, usually consisting of energies and forces obtained from quantum mechanical calculations, such as density functional theory (DFT). Since DFT itself is based on several approximations, MLIPs may inherit systematic errors that lead to discrepancies with experimental data. In this paper, we use a trajectory re-weighting technique to refine DFT pre-trained MLIPs to match the target experimental Extended X-ray Absorption Fine Structure (EXAFS) spectra. EXAFS spectra are sensitive to the local structural environment around an absorbing atom. Thus, refining an MLIP to improve agreement with experimental EXAFS spectra also improves the MLIP prediction of other structural properties that are not directly involved in the refinement process. We combine this re-weighting technique with transfer learning to avoid overfitting to the limited experimental data. The refinement approach demonstrates significant improvement for two MLIPs, one for an established nuclear fuel: uranium dioxide (UO$_2$) and second one for a nuclear fuel candidate: uranium mononitride (UN). We validate the effectiveness of our approach by comparing the results obtained from the original (unrefined) DFT-based MLIP and the EXAFS-refined MLIP across various properties, such as lattice parameters, bulk modulus, heat capacity, point defect energies, elastic constants, phonon dispersion spectra, and diffusion coefficients. An accurate MLIP for nuclear fuels is extremely beneficial as it enables reliable atomistic simulation, which greatly reduces the need for large number of expensive and inherently dangerous experimental nuclear integral tests, traditionally required for the qualification of efficient and resilient fuel candidates.

\vspace{1em} 
\section*{Introduction}

Molecular dynamics (MD) is an essential tool for many advancements in the fields of material science, biophysics, and chemistry~\cite{HOLLINGSWORTH20181129}. However, the quality of MD simulations depends almost entirely on the accuracy of the interatomic potential used to calculate atomic forces. Although classical force fields offer the benefit of computational efficiency for predicting atomic forces, their accuracy is limited by the physics-based functional form and empirical model parameters that are fit to limited experimental data and/or quantum mechanical calculations~\cite{10.1063/5.0011346}. In contrast, quantum mechanical calculations of atomic forces are accurate and transferable to a wide range of systems. However, these calculations are extremely computationally expensive, limiting their applicability to smaller systems.

Machine learning interatomic potentials (MLIPs) are designed to bridge the gap between computational efficiency and accuracy~\cite{PhysRevLett.98.146401, Batzner2022, Ko2023}, while also offering transferability and flexibility to learn many-body interactions. The quality of MLIPs, both in terms of accuracy and robustness, depends almost exclusively on the training data~\cite{doi:10.1021/acs.chemrev.4c00572}. Typically, this training dataset is generated by performing thousands to millions of quantum mechanical calculations to obtain the energies and forces for each atomic configuration~\cite{C6SC05720A}. Active learning procedures are often employed to ensure that the training dataset is minimal in size, diverse in atomic configurations, and adequately covers the relevant regions of the potential energy surface, thus stabilizing MD simulations performed with the resulting MLIP~\cite{10.1063/1.5023802, Sivaraman2020, WILSON2022111330, vanderOord2023, PODRYABINKIN2017171}. Density functional theory (DFT) offers reasonable accuracy and computational feasibility for dataset generation. However, when DFT is not sufficiently accurate, transfer learning can be a powerful technique in which a DFT-trained MLIP is re-trained to a smaller dataset computed with a higher accuracy (and more expensive) quantum mechanical method. While the DFT pre-trained model already captures the broad knowledge of interatomic interactions, the more accurate small dataset can refine the model by fine-tuning certain parameters~\cite{Smith2019_2}. The combination of active learning and transfer learning greatly reduces computational cost and data requirements while achieving higher predictive accuracy. Nonetheless, the accuracy of the resulting MLIP is still limited by the accuracy of the underlying training dataset.

A possible solution to this fundamental limitation is to train directly to experimental data~\cite{Thaler2021, Röcken2024, doi:10.1021/acs.jctc.3c01051}. Classical and coarse-grained potentials are routinely parameterized with experimental data and, thus, tend to outperform MLIPs when tested against similar experimental observables. Leveraging experimental data to systematically construct MLIPs is not as straightforward as training to a quantum mechanical dataset~\cite{D2DD00137C}. Typically, an MLIP is trained using well established techniques, such as gradient descent, in which the atomistic positions are mapped to the corresponding target energies and forces. Such atomistic details are not measurable experimentally. Likewise, the experimental observables cannot be predicted directly from the MLIP. Rather, estimating experimental observables for an MLIP requires performing MD simulations and computing statistical averages of the instantaneous value for the macroscopic property over the MD trajectory. Furthermore, the measured quantities are susceptible to both experimental and simulation fluctuations, making training more challenging.

Training to experimental data has previously been explored using automatic differentiation~\cite{JMLR:v18:17-468,Schoenholz_2021,10.1063/5.0126475}, in which the model parameters are updated at the repetition of forward and backward passes. In the forward pass, an MD simulation is performed and observables are calculated. A loss function based on the difference between the predicted and target MD observables is defined. The gradients of the loss function with respect to the potential energy parameters are calculated in the backward pass. This approach suffers from three major problems: computational inefficiency, memory overflow, and explosive gradients\cite{article1}.

The differential trajectory re-weighting method aims to overcome the challenges of automatic differentiation~\cite{Thaler2021}. In this approach, instead of calculating a new trajectory, a reference trajectory is re-weighted at every update of the MLIP parameters. Moreover, the functional relationship between observables and MLIP parameters established by re-weighting eliminates the calculation of back-propagation gradient at every step. This approach also allows for simultaneous training to both quantum mechanical data and experimental observables~\cite{Röcken2024}. Previous work applying this technique demonstrates that the trained MLIP achieves good agreement with both the experimental observables included in the training as well as observables not included in the training dataset~\cite{Thaler2021,Röcken2024,Lee2024,Han2025,FUCHS2025109512}.

Another approach for refining an MLIP to experimental data is the density alignment method~\cite{gong_predictive_2025}. For this method, an MD simulation with constant temperature and pressure is performed with the pre-trained MLIP. Atomic positions and instantaneous box sizes are stored periodically. The deviation between the pre-trained MLIP simulation density for a given snapshot and the experimental density is included in the loss function. By assuming isothermal compressibility and applying the virial theorem for pressure in terms of the atomic forces, the derivative of density with respect to the MLIP parameters can then be derived with automatic differentiation. Thus, a key advantage of the density alignment method is the ability to apply standard backpropagation to train the MLIP parameters directly to experimental density data. The primary disadvantage of this method is that it is currently limited to density, although similar techniques may be applicable to other properties.

In this work, we use an approach similar to differential trajectory re-weighting where the statistical average for a set of simulation atomistic configurations are re-weighted in accordance with thermodynamic perturbation theory. However, in contrast to differential trajectory re-weighting, the proposed methodology does not require auto-differentiation of the macroscopic observable with respect to the MLIP parameters, rendering the retraining extremely fast and more straightforward. We refine the MLIP indirectly with the experimental spectroscopic observable known as Extended X-ray Absorption Fine Structure (EXAFS) spectra. EXAFS is an ideal property for training because of its high sensitivity to the local structural environment around an absorbing atom. However, our re-weighting technique is amenable to any thermophysical property that can be computed for a given atomic configuration, e.g., the radial distribution function (RDF). To avoid overfitting to a single experimental spectrum, the re-weighting technique is combined with transfer learning.

We apply our refinement method to improve pre-trained MLIPs for an established nuclear fuel UO$_2$ and a nuclear fuel candidate UN. Accurate modeling of these materials can significantly improve mechanistic and fuel performance models, reducing the number of challenging irradiation experiments needed to inform the design and licensing of new reactor systems and fuels. Developing the interatomic potential models for nuclear fuels is particularly important because performing experiments is challenging due to their radioactive nature. Moreover, quantum mechanical calculations are extremely expensive for $f$-electron elements, since the computational cost of traditional Kohn-Sham DFT method scales cubically with the system size~\cite{PhysRev.140.A1133, PhysRev.136.B864}, motivating the need for accurate MLIPs for these materials.

Recently, MLIPs were developed for uranium dioxide (UO$_2$)~\cite{STIPPELL2024100042, PhysRevMaterials.8.025402} and uranium mononitride (UN)~\cite{alzatevargas2024machinelearninginteratomicpotential} using an active learning procedure to generate the training datasets. As the accuracy of an MLIP is limited by the accuracy of the underlying reference data, these previous studies considered several different DFT functionals and values for the Hubbard parameter (+U), which is often added to the exchange correlation functional to account for the strong correlation among uranium 5$f$ electrons. Because standard DFT calculations of UO$_2$ yield a metallic system in the ground state, in contrast to the insulator property known experimentally~\cite{PhysRevB.79.235125, PhysRevB.57.1505}, the UO$_2$ reference data were computed using DFT+U, which correctly captures both the insulator behavior and the antiferromagnetic character of UO$_2$~\cite{STIPPELL2024100042, PhysRevMaterials.8.025402}. By contrast, the UN reference data were computed using standard DFT calculations, as DFT+U yields negative phonon frequencies for UN~\cite{alzatevargas2024machinelearninginteratomicpotential}. 

Both the DFT+U-based UO$_2$ MLIP of Stippell et al.~\cite{STIPPELL2024100042} and the DFT-based UN MLIP of Alzate-Vargas et al.~\cite{alzatevargas2024machinelearninginteratomicpotential} provide a reasonable prediction of point defect energies, elastic constants, and other thermophysical properties over a range of temperatures. However, certain properties are poorly predicted by the respective MLIPs. For example, the UO$_2$ MLIP quantitatively overestimates the lattice parameter by $\approx1.5\%$ compared to the experimental value. This overprediction in lattice parameter is a well-known limitation of the DFT+U method\cite{doi:10.1021/acs.jpcc.2c03804}. To a lesser extent, the UN MLIP underestimates the lattice parameter by $\approx0.5\%$ compared to the experimental value. However, the MLIP lattice parameter is in close agreement with the DFT value\cite{KOCEVSKI2023154241}. Thus, the deviations in lattice parameter with respect to experiment for both the UO$_2$ and UN MLIPs can clearly be attributed to a deficiency in the DFT+U/DFT reference data, and not to a lack of diversity in the training dataset or the MLIP architecture.

In this work, new unrefined MLIPs for UO$_2$ and UN are first trained to the corresponding reference DFT+U\cite{STIPPELL2024100042} or DFT datasets\cite{alzatevargas2024machinelearninginteratomicpotential}. These MLIPs are then refined using our re-weighting technique and synthetic EXAFS data. The improvement in the refined MLIPs is validated by comparing the predicted thermophysical, crystal, energetic, and transport properties with experimental values. In particular, the refined MLIPs demonstrate excellent agreement with the experimental lattice parameter over a wide temperature range, clearly achieving beyond DFT-level accuracy.

\section*{Theory and Methods}

\subsection*{Pre-training the MLIP}

The unrefined MLIP for UO$_2$ was pre-trained on a DFT+U dataset consisting of 3365 structures\cite{STIPPELL2024100042}, while the unrefined MLIP for UN was pre-trained on a DFT dataset consisting of 12336 structures\cite{alzatevargas2024machinelearninginteratomicpotential}. See the literature for details regarding the DFT+U/DFT calculations (e.g., functional, Hubbard parameter).

\begin{figure}
    \centering
    \includegraphics[width=0.95\linewidth]{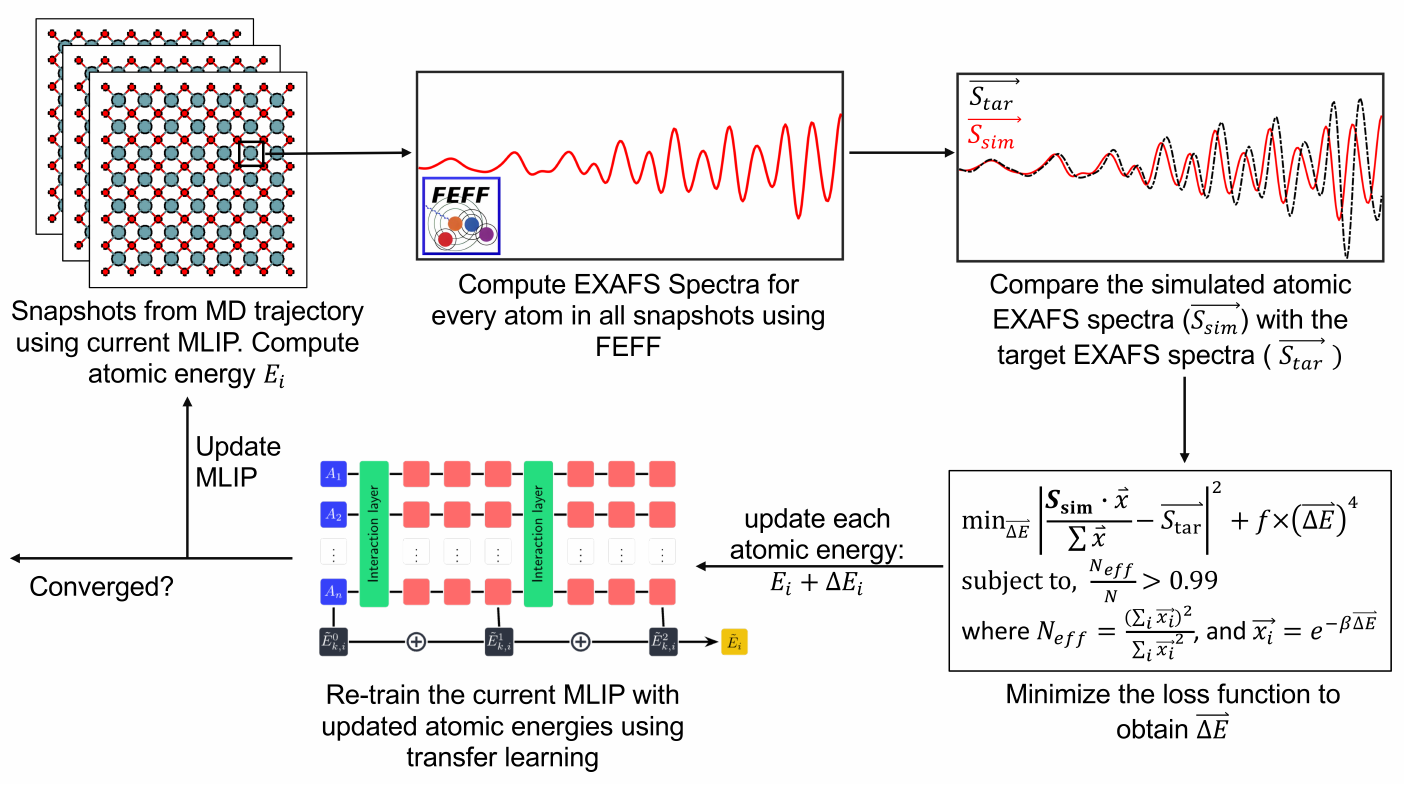}
    \caption{Flowchart for the MLIP refinement process. In this process, the current MLIP for a given iteration is used to generate the MD trajectory. The EXAFS spectra for every atom are calculated for a subset of snapshots from the full trajectory. The atomic energy shifts $(\Delta E)$ are determined by minimizing the loss function, which depends on the difference between the re-weighted simulated EXAFS spectra and the target EXAFS spectra (bottom right box). To avoid large changes in the MLIP between iterations, a regularization term is included in the loss function to penalize large values of $\Delta E_i$. The MLIP is then re-trained with the updated atomic energies, $E_i + \Delta E_i$, where transfer learning is employed to avoid overfitting. At every iteration, the agreement between simulated and target EXAFS spectra improves.}
    \label{fig:flowchart}
\end{figure}

Both MLIPs are trained with the Hierarchical Interacting Particle Neural Network (HIP-NN)~\cite{10.1063/1.5011181}, a message-passing graph neural network architecture (see Figure~\ref{fig:flowchart}). HIP-NN receives as inputs the element types and interatomic distances within the cutoff radius. These inputs are converted into a feature vector ($z_{i,a}$) of the local chemical environment around atom $i$. The feature vector for the input layer is simply the one-hot encoding representation of atom $i$ with length $a$ equal to the number of element types, while the feature vector for all subsequent layers is of length $a$ for $a=1\dots n_{\mathrm{features}}$. The feature vector is passed through a series of interaction blocks, consisting of an interaction layer followed by $n_{\rm layers}$ atomic layers. In the interaction layer, the feature vector of the central atom $i$ is ``mixed'' with the feature vectors for all of the neighboring atoms within a given cutoff radius. 

HIP-NN returns as output the so-called atomic energy, i.e., the contribution to the total energy from a single atom. Although the atomic energy is not a physical property that can be computed with quantum mechanical methods, the atomic energy is a useful construct in classical force fields and MLIPs. The total energy $(E)$ of an atomic configuration is simply the sum of atomic energies $(E_i)$ for all atoms $(n_{\text{atom}})$ in the system
\begin{equation}
    E = \sum_{i=1}^{n_{\text{atom}}} E_i
\end{equation}

Compared to other MLIP architectures, a unique aspect of HIP-NN is that it predicts the atomic energy as a sum of $n$ hierarchical energies such that
\begin{equation}
    E_i = \sum_{n=0}^{n_{\text{int}}-1} E^n_i
\end{equation}
where $E_i^n$ is the $n^{\rm th}$ hierarchical energy for atom $i$, and $n_{\text{int}}$ is the number of interaction blocks. Inspired by the many-body expansion, the hierarchical energy $E_i^n$ represents (n+1)-body interactions. Thus, $E_i^n$ should decay with increasing $n$. To enforce this physical constraint, a regularization term is included in the loss function that helps ensure $E_i^{n+1} < E_i^n$ (see Supporting Information).

The HIP-NN hyperparameters for the UO$_2$ MLIP are tuned to minimize the energy and force RMSE values (see Tables S1, S2, and S3). A comparison is made between different loss functions by evaluating the bulk modulus for UO$_2$ obtained using the resulting MLIPs (see Table S4 and Figure S1). The selected hyperparameters for the UO$_2$ MLIP are included in the Supporting Information. For UN, the same MLIP hyperparameters and loss function reported in previous work are used\cite{alzatevargas2024machinelearninginteratomicpotential}.

\subsection*{Computing the target EXAFS spectra}

Extended X-ray Absorption Fine Structure (EXAFS) spectrum is an extremely valuable target experimental observable due to its high sensitivity to the local structural environment around an absorbing atom\cite{article2,Palmer1996,HIGGINBOTHAM200944,Price_2013}. In EXAFS measurements, an excited electron is absorbed by a material, which causes core-orbital electrons to be ejected. The ejected electrons are observed as an outgoing photo-electron wave, which is back-scattered against the neighboring atoms. The outgoing and back-scattered waves result in an interference pattern, manifested as oscillations in the EXAFS spectra\cite{RevModPhys.72.621}. Neighboring atoms at different positions with respect to the absorbing atom induce a different interference pattern. Hence, the EXAFS spectrum is a function of near-neighbor distances, coordination number, and bond-distance fluctuations.

Within the single-scattering approximation, the EXAFS function $\chi$ is defined as

\begin{equation}
\chi(k) = \sum_{j=1}^{j} A_j(k) \sin(2kR_j + \phi_j(k))
\label{exafsfunction}
\end{equation}
where $k$ is the wavevector, $j$ refers to the coordination shell around the absorbing atom, $R_j$ is the distance between the absorbing atom and an atom in the $j^{\rm th}$ coordination shell, and $\phi_j(k)$ is the phase shift experienced by photo-electron during scattering. The oscillations in the EXAFS spectra are governed by the sine function. Since these oscillations depend on $R_j$, refining an MLIP to match the experimental EXAFS oscillations should improve the interatomic distances and the geometric arrangement around the absorbing atom. $A_j(k)$ is the back-scattering amplitude described as follows:

\begin{equation}
A_j(k) = \frac{S_0^2 e^{-\frac{2R_j}{\lambda}}}{kR_j^2} \times e^{-2\sigma_j^2 k^2} \times N_j F_j(k)
\label{amplitude}
\end{equation}
where the first term indicates the damping of the signal due to the loss of photo-electron energy, where $S_0^2$ is the so-called amplitude reduction factor and $\lambda$ is the mean free path. The second term describes static and thermal disorders in material caused by fluctuations in the interatomic distance, where $\sigma_j^2$ is the Debye-Waller factor. The third term accounts for the scattering power of the coordination shell $j$, where $N_j$ is the number of neighboring atoms in the $j^{th}$ shell, and $F_j(k)$ is the scattering power of the neighbor $j$.

Retraining an MLIP to experimental EXAFS spectra requires a spectrum for all elements in the system. Preliminary attempts to retrain our UO$_2$ MLIP with an EXAFS spectrum for only U atoms (see Figure S2) resulted in significant degradation of the MLIP (see Figure S3 and the next section for a detailed explanation). Unfortunately, for both UO$_2$ and UN, experimental EXAFS spectra are only available for U atoms. Therefore, we develop synthetic EXAFS spectra for each element, i.e., a separate spectrum for U and O in UO$_2$, and for U and N in UN. These synthetic spectra serve as target macroscopic observables in the refinement procedure of their corresponding MLIPs. Each synthetic spectrum is generated for a perfect crystal modeled with the experimental lattice parameter (see Figure S4). The experimental lattice parameter is used to ensure accurate interatomic distances and coordination shells for the perfect crystal and, consequently, to improve agreement between the synthetic EXAFS spectra and the experimental spectra. Generating synthetic spectra would not be necessary for systems where experimental EXAFS spectra are available for all elements. EXAFS spectra are needed for all elements even when using synthetic spectra, as evidenced by the sporadic behavior of the refined MLIP at high temperatures if the refinement process is performed with synthetic spectra for either U or O (see Figure S5).

A comparison between uranium synthetic and experimental EXAFS spectra in UO$_2$ at 77 K is shown in Figure S6. The peak positions of the EXAFS spectra along the wavevector $k$ align well. Thus, the oscillations of the synthetic EXAFS spectra are consistent with the experimental spectra. However, the perfect crystal does not consider thermal vibrations or non-crystallinity present in the experimental measurements. Thermal effects dampen the spectral peaks, as the EXAFS amplitude is directly proportional to the Debye-Waller factor (second term in Equation~\ref{amplitude}), which decreases exponentially with increasing mean square displacement of atoms $\sigma_j^2$~\cite{PhysRevB.76.014301}. For this reason, the peaks in the experimental EXAFS spectra are dampened compared to the perfect crystal (see Figure S6). To mitigate thermal effects, the MLIP refinement procedure is employed to match the synthetic EXAFS spectra at low temperatures 75~K for UO$_2$ and 50~K for UN. Preliminary attempts to refine the MLIP with experimental EXAFS spectra for U at higher temperatures (300~K) resulted in an unstable MLIP (see Figures S7 and S8).

In general, the refinement process does not necessarily require synthetic target spectra, as long as experimental data are available for all elements. We provide a simple example where experimental spectra are available for pure aluminum (see Appendix). In this example, the pre-trained MLIP is refined using RDF data at two temperatures, demonstrating that the refinement process is not limited to EXAFS data and is applicable to multiple temperatures.

\subsection*{Refining the MLIP}

Figure~\ref{fig:flowchart} shows the detailed MLIP refinement workflow. During this process, the MLIP pre-trained to the DFT/DFT+U dataset is iteratively re-trained to improve agreement with the target EXAFS spectra. Rather than re-training the unrefined MLIP directly to EXAFS data, our refinement methodology utilizes the EXAFS spectra to generate a new refinement dataset consisting of shifted atomic energies.

The refinement dataset is produced using the trajectory re-weighting technique in a multi-step process. The first step is to perform a set of MD simulations with the current MLIP. In the $0^{\rm th}$ iteration, the current MLIP is the unrefined model pre-trained to DFT/DFT+U data. Snapshots, i.e., atomic configurations, from the MD trajectory are selected at equal time intervals. The atomic energies ($E_{i}$) for each snapshot are computed using the current MLIP. In the second step, EXAFS spectra are obtained for every atom in each snapshot. Third, these simulated EXAFS spectra are re-weighted to better match the target EXAFS spectra by minimizing a loss function (see Equation~\ref{lossfun}). The $i^{\rm th}$ component in the optimal re-weighting vector ($\vec{w}$) is used to calculate atomic energy shifts $(\Delta E_{i})$, see Equation~\ref{eq:w8} and Equation~\ref{eq:rew8}. The refinement dataset consists of atomic configurations and shifted atomic energies $(E_{i} + \Delta E_{i})$ for every atom in each snapshot. Because $\Delta E_{i}$ is computed only for elements with target EXAFS spectra, the atomic energies for elements without target EXAFS spectra would remain unshifted ($\Delta E_{i}) = 0$), resulting in a data inconsistency when EXAFS spectra are not available for all elements.

Notably,because the original DFT/DFT+U forces are not compatible with the shifted atomic energies, the refinement dataset does not contain atomic forces. There are at least two key reasons why training with forces typically results in more robust MLIPs than simply training to total energy~\cite{smith2020simpleefficientalgorithmstraining}. First, whereas total energy is a global property, atomic forces are a local property and, thus, are more rich in information content regarding the interactions governing a given atomic environment. Second, while each snapshot is assigned a single total energy, every atom contains a 3-dimensional vector of forces, greatly increasing the shear amount of data compared to a dataset consisting of solely total energies. Training to atomic energies is, thus, more similar to training to atomic forces because atomic energies are a local property and every atom is assigned an atomic energy.

The shifts in atomic energy are obtained by minimizing the loss function:
\begin{equation}
\label{lossfun}
\text{min}_{\Delta{\vec{E}}} \left| \bm{S}_{\rm rw} - {\vec{S}_{\rm tar}}\right|^2 + {f}\times(\Delta{\vec{E}})^4
\end{equation}
where $\Delta \vec{E}$ is a vector for the shifts in atomic energies, $\vec{S}_{\rm tar}$ is the target EXAFS spectrum, and $\bm{S}_{\rm rw}$ is a matrix of re-weighted spectra. The re-weighted spectra are defined as:
\begin{equation}
\label{eq:Srw8}
\bm{S}_{\rm rw} = {\bm{S}_{\rm sim} \cdot \vec{w}}
\end{equation}
where $\bm{S}_{\rm sim}$ is a matrix of simulation spectra and the re-weighting vector, $\vec{w}$, assigns each simulation spectrum a weight between zero and one. Inspired by thermodynamic perturbation theory~\cite{thermodynamic_perturbation}, the re-weighting vector is defined as:
\begin{equation}
\label{eq:w8}
\vec{w} = \frac{\vec{x}} {\sum {\vec{x}}}
\end{equation}
where each component in vector $\vec{x}$ is given by the Boltzmann factor, i.e., 
\begin{equation} \label{eq:rew8}
    \vec{x} = e^{-\beta{\Delta}\vec{E}}
\end{equation}
where $\beta\equiv\frac{1}{k_{\rm B}T}$, $k_{\rm B}$ is the Boltzmann constant, and $T$ is the simulation temperature. Because the Boltzmann factor depends exponentially on $-\Delta E$, configurations with more negative energy shifts are assigned weights closer to one, whereas configurations with less negative energy shifts are assigned weights closer to zero.

Thermodynamic perturbation theory, and by extension our re-weighting technique, is reliable only for small updates to the MLIP, or shifts to the atomic energy~\cite{messerly_configuration-sampling-based_2018,doi:10.1021/acs.jced.8b01232}. Large shifts in atomic energy lead to a strong imbalance in the re-weighting vector, where only a few configurations are assigned non-negligible weights. An imbalanced re-weighting vector results in poor statistical averaging and, thus, poor estimates of the re-weighted spectrum. Large shifts in atomic energy can also cause the refined MLIP to completely forget the qualitative landscape of the potential energy surface from the DFT/DFT+U dataset. For these reasons, a regularization term is added to the loss function (2nd term in Equation~\ref{lossfun}) to ensure that the absolute value of the atomic energy shifts are closer to zero. However, a balance in $\Delta \vec{E}$ is necessary to ensure reliable estimates of the re-weighted spectra while also significantly improving the MLIP. To penalize large absolute values in $\Delta \vec E$ while also allowing for meaningful shifts in the atomic energy, the regularization factor $f$ is chosen to be the smallest value that maintains the following inequality constraint: 
\begin{equation}
\frac{N_{\text{eff}}}{N} > 0.99
\end{equation}
where $N$ is the sum of the total number of atoms in all of the snapshots in the refinement dataset and $N_{\rm eff}$ is the number of effective samples, i.e., the number of configurations that contribute meaningfully to the re-weighted EXAFS spectra, defined by:
\begin{equation}
N_{\text{eff}} = \frac{(\sum_{i=1}^{N}\vec{x}_{i})^2}{\sum_{i=1}^{N} \vec{x}_{i}^2}
\end{equation}
where $N_{\text{eff}} = N$ for $\Delta \vec{E} = \vec 0$ and $N_{\text{eff}} \approx 1$ for strongly imbalanced $\Delta \vec{E}$. The loss function (Equation~\ref{lossfun}) is minimized using the optimization algorithm LM-BFGS (Limited-memory Broyden–Fletcher–Goldfarb–Shanno)~\cite{Liu1989}.

The obtained shifted atomic energies are used to fine-tune the pre-trained MLIP parameters. In our refinement method, the MLIP is not trained directly to the target macroscopic observable. Instead, the MLIP learns indirectly from the EXAFS spectra by training to the shifted atomic energies, which are obtained through the EXAFS re-weighting optimization (Equation~\ref{lossfun}). Thus, although similar in nature, our refinement method differs from automatic differentiation and differential trajectory re-weighting methods that refine MLIPs directly to macroscopic experimental observables. One key benefit of our approach is that standard ML techniques can be employed when re-training the MLIP to the refinement dataset. For example, we implement transfer learning to avoid overfitting to the refinement dataset and catastrophic forgetting of the DFT/DFT+U training dataset (see~\nameref{sec:simulationDetails}). 

\section*{Results}

To achieve beyond DFT-level accuracy, we refine the pre-trained MLIP of uranium dioxide and uranium mononitride to match target synthetic EXAFS spectra. The unrefined MLIPs are pre-trained to DFT/DFT+U total energy and atomic forces, while the refined MLIPs are re-trained with transfer learning to the shifted atomic energies. Although only atomic-scale quantities are involved in the initial training and refinement process, the refined MLIP predicts improved macroscopic properties beyond EXAFS, that are not explicitly targeted in the training process. To validate the improvement and robustness of the refined UO$_2$ and UN MLIPs, we calculate some relevant properties and compare the MLIP predictions with available experimental data, for temperature-dependent thermophysical properties (lattice parameters, bulk modulus and heat capacity), crystal properties (elastic constants and phonon dispersion), energetic properties (point defect energies), and transport properties (diffusion coefficients).

We first verify the accuracy of the UO$_2$ and UN unrefined MLIPs pre-trained to the respective DFT+U and DFT datasets. Table~\ref{tab:rmse} reports the energy-per-atom and force root-mean-square error (RMSE) between the unrefined MLIP-predicted values and the corresponding DFT+U/DFT reference values for UO$_2$ and UN. The energy RMSE values are similar for both MLIPs and less than 10 meV/atom. The force RMSE for the UO$_2$ MLIP is approximately 0.435 eV/\AA, while the force RMSE for the UN MLIP is around 0.136 eV/\AA. The relatively larger force RMSE value of UO$_2$ is a consequence of the reduced DFT+U dataset consisting of more challenging atomic configurations, with fewer near-equilibrium low-energy configurations. For comparison, the RMSEs for UN (3.9 meV/atom for energy and 0.136 eV/\AA~for forces) are only slightly higher than those reported previously (2.48 meV/atom for energy and 0.113 eV/\AA~for forces) for the same HIP-NN architecture trained on the same dataset~\cite{alzatevargas2024machinelearninginteratomicpotential}. Furthermore, the RMSEs for both UO$_2$ and UN are of the same magnitude to the reported values for a different MLIP architecture (4.5 meV/atom for energy and 0.380 eV/\AA~for forces for UO$_2$ and 2.3 meV/atom for energy and 0.072 eV/\AA~for forces for UN), demonstrating that our model achieves the expected level of accuracy on these training datasets~\cite{STIPPELL2024100042,alzatevargas2024machinelearninginteratomicpotential}.

\begin{table}[ht]
\centering
\caption{The root-mean-square error (RMSE) between the reference DFT+U/DFT energies and forces and those predicted by the unrefined MLIPs for UO$_2$ and UN using a held-out test dataset.}
\label{tab:rmse}
\begin{tabular}{@{}lcccccc@{}}
\toprule
Material & Energy (meV/atom) & Force (eV/\AA) \\ 
\midrule
 UO$_2$  & 6.5 & 0.435  \\
 UN  & 3.9 & 0.136  \\
\bottomrule
\end{tabular}
\end{table}

Because the refined MLIPs are trained to shifted atomic energies, the RMSEs for the refined MLIPs with respect to the DFT+U/DFT energies are significantly higher (compare Figures S9(a) and S10(a) with Figures S9(c) and S10(c)). However, the energy errors are almost entirely positive for UO$_2$ and negative for UN, demonstrating that the energy shifts are systematic and internally consistent. Furthermore, although atomic forces were not utilized during retraining, the refined MLIPs predict the DFT+U/DFT forces with nearly the same accuracy as the unrefined MLIPs (compare Figures S9(b) and S10(b) with Figures S9(d) and S10(d)). Some increase in force errors is expected, as the DFT+U/DFT-level forces are not consistent with the shifted atomic energies. However, the relatively small increase in force errors demonstrates that the refined MLIPs did not completely ``forget'' the fundamental physics contained in the DFT+U/DFT training datasets.

\subsection*{Uranium Dioxide}

Since the refinement process is performed with respect to the synthetic target EXAFS spectra for both U and O, we verify that the refined MLIP predicts spectra in closer agreement with the experimental EXAFS spectrum for U. Because refinement is performed at a single temperature of 75~K, we also verify the improvement for EXAFS spectrum prediction at 300~K.

\begin{figure*}[t]
    \centering
    \includegraphics[width=0.6\linewidth]{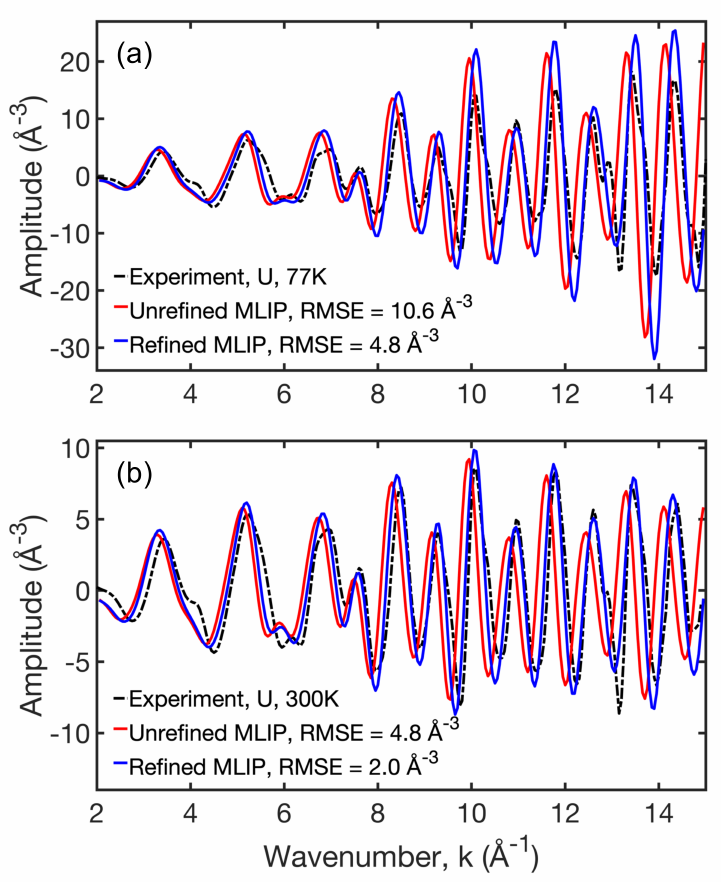}
    \caption{Comparison of experimental EXAFS spectra of U in UO$_2$ with the simulated EXAFS spectra computed with unrefined MLIP and refined MLIP at (a) 77 K and (b) 300 K. While the refinement process is performed at a single temperature (75 K), the refined MLIP predicts improved EXAFS spectra even at higher temperatures. To ensure an unbiased comparison with experiments, accurate SCF method was used to generate the EXAFS spectra corresponding to unrefined and refined MLIP. The simulated EXAFS spectra is obtained by averaging over all U atoms in a single $4 \times 4 \times 4$ supercell of UO$_2$ from molecular dynamics trajectory at respective temperatures.}
    \label{fig:exptcomparison}
\end{figure*}

Figure~\ref{fig:exptcomparison} presents a comparison of the experimental EXAFS spectrum for uranium in UO$_2$ with those computed using unrefined and refined MLIPs at 77~K (a) and at 300~K (b). At both temperatures, the spectral peaks shift towards higher k-values, aligning more closely with the experimental data. At 77~K, the peak amplitude of the refined MLIP increases slightly compared to the unrefined MLIP. However, both the unrefined and refined spectra have an enhanced peak amplitude compared to the experimental spectra. At 300~K, we observe better agreement of the peak amplitude with the experimental spectrum. 

Even though the amplitudes for the synthetic perfect crystal spectrum are significantly larger than the corresponding ones in the experimental spectrum (see Figure S6), the refined MLIP spectral peak amplitudes are in reasonable agreement with experiment. Overfitting to the synthetic spectra could induce unusually strong interatomic forces, reducing thermal vibrations and producing overly large peak amplitudes. Careful transfer learning (see~\nameref{sec:simulationDetails}) helps avoid overfitting and catastrophic forgetting of the initial DFT+U training dataset. Figure S11 also shows systematic and gradual improvement in the MLIP-generated EXAFS spectra with respect to the synthetic target spectra, suggesting that the EXAFS refinement approach converges smoothly.

Table~\ref{tab:elasticuo2} reports the UO$_2$ elastic constants at 296~K for both MLIPs (unrefined and refined) compared to the experimental values at 296 K~\cite{10.1063/1.322438}. The refined MLIP shows a significant improvement in the underestimated $C_{11}$ and $C_{44}$ constants, while still overestimating the $C_{12}$ constant. 

\begin{table}[h!]
\centering
\caption{Elastic constants for UO$_2$, calculated with unrefined and refined MLIPs at 296 K, compared with the experimental values at 296~K from Ref.~\cite{10.1063/1.322438}.}
\label{tab:elasticuo2}
\begin{tabular}{ccccc}
\toprule
Elastic constants & $C_{11}$ (GPa) & $C_{12}$ (GPa) & $C_{44}$ (GPa) \\ 
\midrule
 Experiment & 389.3 $\pm$ 1.7 & 118.7 $\pm$ 1.7 & 59.7 $\pm$ 0.3 \\
 Unrefined MLIP & 346.02 & 140.18 & 44.68 \\
 Refined MLIP & 367.10 & 143.82 & 49.20 \\
\bottomrule
\end{tabular}
\end{table}

\begin{figure*}[ht!]
    \centering
    \includegraphics[width=1.0\linewidth]{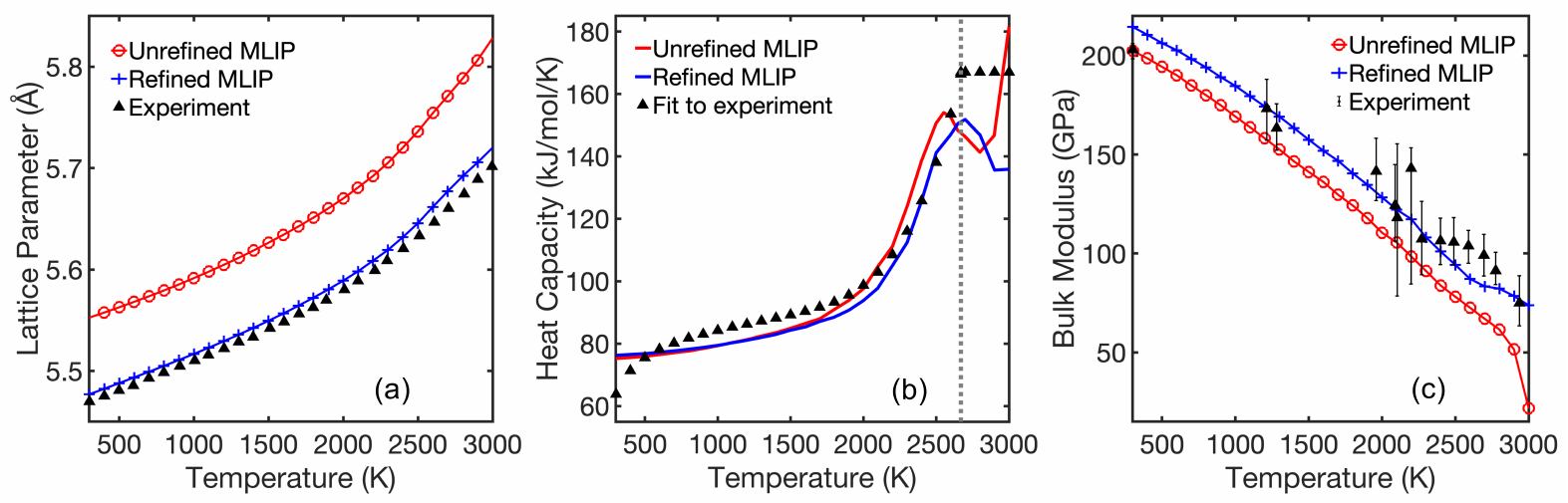}
    \caption{Temperature dependence of thermophysical properties of UO$_2$, (a) lattice parameter, (b) heat capacity, and (c) bulk modulus over a range of temperature from 300 K to 3000 K. Predictions obtained using refined and unrefined MLIPs are compared with the experimental values. In (b), the dashed line is plotted at T=2670 K corresponding to the Bredig temperature. The experimental lattice parameters are from Ref.~\cite{international2006iaea, FINK20001}, the polynomial fit to experimental heat capacity data is from Fink et al. \cite{FINK20001}, and the experimental bulk modulus data is from Hutchings et al.~\cite{F29878301083}.}
    \label{fig:figg4}
\end{figure*}

Figure~\ref{fig:figg4} shows the predicted temperature-dependent (a) lattice parameter, (b) heat capacity, and (c) bulk modulus using the unrefined and refined MLIPs in comparison to experimental data. The systematic and gradual improvement of these properties at each refinement iteration is shown in Figure S12. As shown in Figure~\ref{fig:figg4}(a) and (c), the most significant improvements achieved through MLIP refinement are the enhanced accuracy of lattice parameter and bulk modulus with respect to temperature. The percent improvements in lattice parameter and bulk modulus of approximately 1.5\% and 15\%, respectively, are comparable with those achieved when refining a DFT-trained MLIP for titanium using differentiable trajectory reweighting~\cite{Röcken2024}.

Because EXAFS spectra depend strongly on the local atomic environment and the interatomic distances around the absorbing atom, the improvement in lattice parameter can be attributed to a better agreement with the target EXAFS spectra, which was generated with the experimental lattice parameter. However, the MLIP refinement process does not train directly to lattice parameter data. Thus, the lattice parameter temperature dependence represents an out-of-sample test case. Remarkably, the refined MLIP produces improved lattice parameters over a wide range of temperatures from 300~K to 3000~K, even though the refinement is performed at a single low temperature (75~K). Previous work also reports similar improvement in lattice parameter over a wide range of temperatures when refining with experimental data at a single temperature using differentiable trajectory reweighting~\cite{Röcken2024}. Figure~\ref{fig:figg4}(a) demonstrates that the steady overprediction of lattice parameter given by the unrefined MLIP gradually deflects above 2000~K. This deflection is also improved in the lattice parameters predicted by the refined MLIP, suggesting that the refined MLIP is more stable at higher temperatures. This improvement can be more clearly observed from the thermal expansion plot at higher temperatures, from 2500~K to 3000~K shown in Figure S13. Such consistent improvement over a range of temperature is not observed when the re-training is done without freezing certain HIP-NN layers (Figure S14 and Figure S15), which implies that freezing some layers during transfer learning eliminates possible overfitting to the small refinement dataset.

Figure~\ref{fig:figg4}(b) shows the temperature-dependent heat capacity for unrefined and refined MLIPs compared to the experimental results. The increase in heat capacity with temperature for UO$_2$ has been attributed to different factors in different temperature ranges. At low temperatures up to 1000~K, the strongest contribution to heat capacity are harmonic lattice vibrations, however between 1000~K to 1500~K, there is an increase in anharmonicity. The sharp increase of heat capacity observed from 1500~K up to $\approx$2670~K, has been linked to the formation of lattice and electronic defects~\cite{international2006iaea}. An abrupt transition in heat capacity, named Bredig temperature is observed experimentally at 2670~K and is considered a result of the saturation of defect concentration~\cite{internationalsurl'étudedestransformationscristallinesàhautetempératureau-dessusde2000K1972}. While both refined and unrefined MLIPs reproduce this transition phenomenon, the refined MLIP predicts a slightly more accurate transition temperature compared to the unrefined MLIP, again indicative of a better description of UO$_2$ at high temperatures.

Figure~\ref{fig:figg4}(c) shows improvement in the refined MLIP predictions for bulk modulus as a function of temperature. The predicted bulk modulus using the unrefined MLIP is underestimated with respect to the experimental data of Hutchings et al.~\cite{F29878301083}, except at lower temperatures, where the MLIP matches the experimental value at 296~K. In contrast, the refined MLIP predictions are consistent over the entire temperature range and in closer agreement--even at high temperatures--with the experimental values. Overall, the refined MLIP predicts higher bulk modulus compared to the unrefined MLIP, consistent with higher EXAFS peak amplitudes and a stiffer response to volumetric compression.

\begin{figure*}[ht!]
    \centering
    \includegraphics[width=0.9\linewidth]{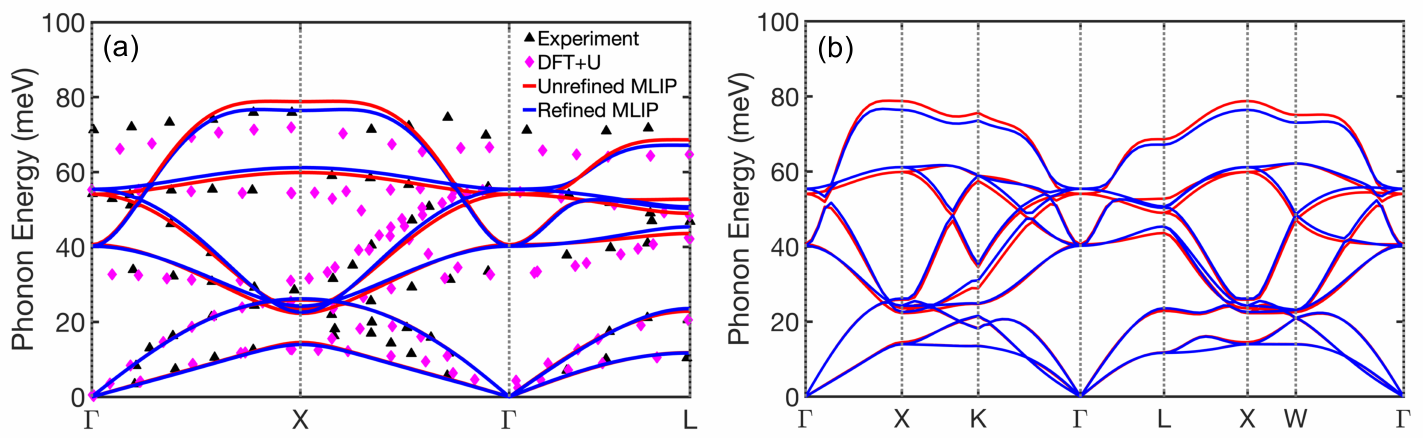}
    \caption{Phonon dispersion spectra for UO$_2$ calculated using unrefined and refined MLIPs, for a shorter path (a) $\Gamma \to X \to \Gamma \to L$, and an extended path (b) $\Gamma \to X \to K \to \Gamma \to L \to X \to W \to \Gamma$. The experimental inelastic neutron scattering data at 300 K in (a) is obtained from Peng et al. \cite{PhysRevLett.110.157401}, and DFT+U calculated dispersion spectra from \cite{phonon_dftu}.}
    \label{fig:figg5}
\end{figure*}

Figure~\ref{fig:figg5} shows that the phonon dispersion spectra are similar for both the unrefined and refined MLIP. This similarity reflects that UO$_2$ lattice vibrations are not affected significantly after refinement. Since, the synthetic target spectra is computed for a perfect crystal, overfitting to the synthetic target spectra during refinement would lead to stiffer vibrations and degrade phonon dispersions. However, the phonon spectra remain relatively unchanged, demonstrating that the refinement process improves the structural properties while not causing the refined MLIP to forget lattice vibration contributions contained in the DFT+U original training dataset.

\begin{table}[ht!]
\centering
\caption{Point defect energies for UO$_2$ predicted using the unrefined and refined MLIPs and compared with the DFT+U literature values from Refs.~\cite{DORADO2010103,Vathonne_2014}, and experimental values from Matzke et al.~\cite{F29878301121}.}
\label{tab:defectUO2}
\begin{tabular}{@{}lcccccc@{}}
\toprule
Defect energies (eV) & Schottky defect & Frenkel Pair (U) & Frenkel Pair (O) \\ 
\midrule
 Experiment & 6.0-7.0 & 9.5 & 3.0-4.0 \\
 DFT+U & 4.2-11.8 & 9.1-16.5 & 2.4-7.0 \\
 Unrefined MLIP & 6.56 & 10.60 & 4.99 \\
 Refined MLIP & 6.88 & 11.24 & 4.90 \\
\bottomrule
\end{tabular}
\end{table}

Nuclear fuels are continuously exposed to neutron irradiation, leading to the formation of vacancies, interstitial and dislocation defects. Defect energies are critical properties, as they can influence mechanical degradation and ultimately fuel operation lifetime. The unrefined MLIP yields point defect energies that align well with DFT+U results~\cite{DORADO2010103,Vathonne_2014}. Since the refinement process shifts the atomic energies in the MLIP, it is essential to ensure that point defect predictions remain accurate. Table~\ref{tab:defectUO2} reports the point defect energies obtained from both the unrefined and refined MLIPs, compared to experimental and DFT+U literature values~\cite{DORADO2010103, Vathonne_2014,F29878301121}. The Schottky defect energy predicted by both MLIPs falls within the experimental range. Similarly the uranium and oxygen Frenkel pair defect energies are within the DFT+U range, although they are slightly overestimated relative to experimental data. Considering the broad uncertainty in both experimental values and DFT+U calculations, the point defect energies calculated from both the unrefined and refined MLIPs are considered reasonable. This confirms that the refinement process does not negatively affect defect energies.

\begin{figure}[ht!]
    \centering
    \includegraphics[width=0.5\linewidth]{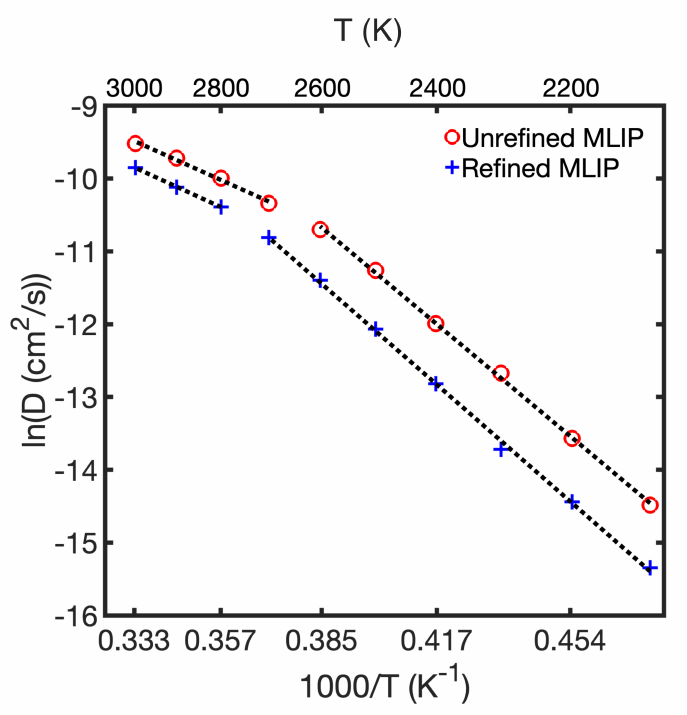}
    \caption{Oxygen diffusion coefficients in UO$_2$ $log(D)$ as a function of inverse temperature (1000/T) (K$^{-1}$) calculated with unrefined and refined MLIPs at temperatures between 2100~K and 3000~K. Linear fits of Arrhenius equation for two regimes are shown as dashed lines.}
    \label{fig:diff_coeff}
\end{figure}

Subsequent validation of transport properties is important when training to only static structural data, such as EXAFS spectra. For this reason, we compare the diffusion coefficient $(D)$ for oxygen in UO$_2$ predicted with the unrefined and refined MLIPs. The simulation methods used to calculate these values are provided in the ~\nameref{sec:simulationDetails} section and Supporting Information (see Figure S16). Figure~\ref{fig:diff_coeff} provides the oxygen diffusion coefficients as a function of inverse temperature from 2100~K to 3000~K. The refinement process increases the diffusion coefficient values approximately by an order of magnitude. For better comparison with experimental data, the activation energies for oxygen diffusion in UO$_2$ are calculated using an Arrhenius fit to the diffusion coefficient. Figure~\ref{fig:diff_coeff} shows that two distinct linear regimes exist. However, the transition temperature that separates these low and high regimes differ between the unrefined MLIP and refined MLIP by 100~K, with the refined temperature of $\approx$2700~K matching better the experimental Bredig temperature of 2670~K.

Table~\ref{tab:activationuo2} reports the activation energy for low- and high-temperature regimes obtained from oxygen diffusion in UO$_2$ and calculated with the unrefined and refined MLIPs. In the high-temperature regime, the activation energy agrees closely with the experimental values~\cite{Kupryazhkin2004}. The activation energy in the low-temperature regime is overestimated for both unrefined and refined MLIPs. Notably, the re-training has minimal impact on the activation energy in both ranges of temperature, demonstrating that the MLIP has retained valuable transport information from the DFT+U dataset.

\begin{table}[ht!]
\centering
\caption{Activation energy for oxygen diffusion in UO$_2$ compared with the experimental values~\cite{Kupryazhkin2004}. The low temperature ranges are: 1800 K to 2600 K for experiment, 2100 K to 2600 K for the unrefined MLIP, and 2100 K to 2700 K for the refined MLIP. The high temperature ranges are: 2600 K to 3100 K for experiment, 2700 K to 3000 K for the unrefined MLIP, and 2800 K to 3000 K for the refined MLIP.}
\label{tab:activationuo2}
\begin{tabular}{@{}lcccccc@{}}
\toprule
Activation energy (eV) & Low temperature & High temperature \\ 
\midrule
 Experiment & 2.6 $\pm$ 0.2 & 1.88 $\pm$ 0.13\\
 Unrefined MLIP & 3.58 & 1.92 \\
 Refined MLIP & 3.73 & 1.95 \\
\bottomrule
\end{tabular}
\end{table}

\subsection*{Uranium Mononitride}

To further validate our methodology, we refine an MLIP for uranium mononitride that was pre-trained to DFT data~\cite{alzatevargas2024machinelearninginteratomicpotential}. In contrast to UO$_2$, DFT+U calculations were not required to properly describe the magnetic and metallic character of UN. Therefore, the unrefined UN MLIP predictions do not differ as strongly from experiment, especially for the lattice parameter. Consequently, unlike UO$_2$, the EXAFS peak positions obtained from the unrefined MLIP already align well with the synthetic target EXAFS spectra for UN (see Figure S17). Although less improvement is needed for UN system than for UO$_2$, this test case demonstrates how it is still important to carefully refine MLIPs without depreciating their accuracy for other properties that are already predicted well from the DFT training dataset.

\begin{figure}[ht!]
    \centering
    \includegraphics[width=0.55\linewidth]{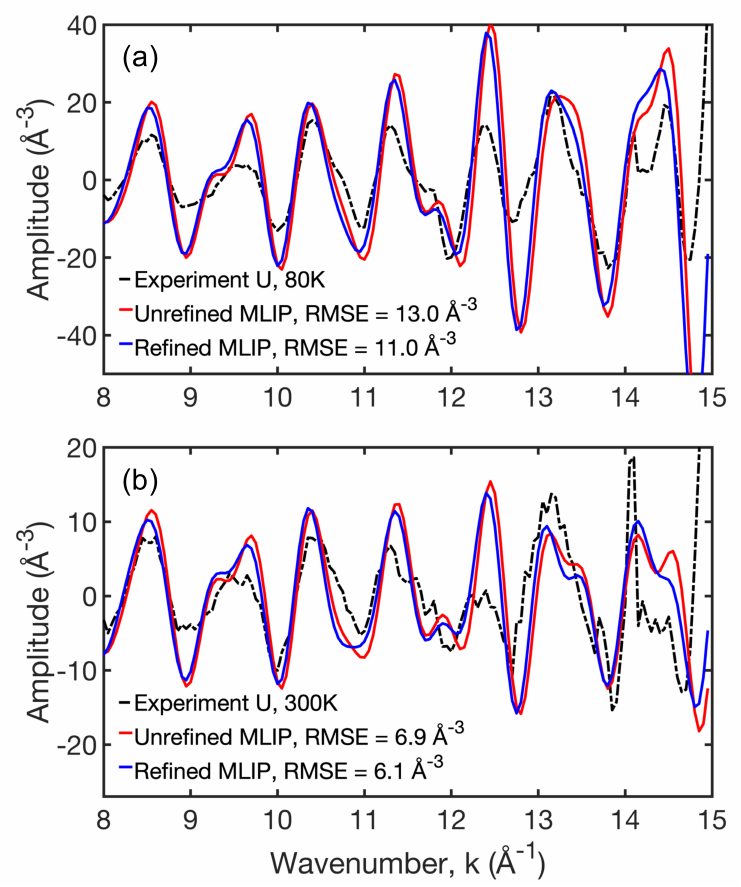}
    \caption{Comparison between experimental EXAFS spectra of uranium atoms in UN and the simulated EXAFS spectra calculated with unrefined and refined MLIPs at (a) 80~K and (b) 300~K. Even as the refinement process for UN MLIP is done at a single low temperature 50~K, the MLIP predicts improved EXAFS spectra at higher temperatures. To ensure an unbiased comparison with experiments, accurate SCF method was used to generate the EXAFS corresponding to unrefined and refined MLIP. The simulated EXAFS spectra is obtained by averaging over all U atoms in a single $5 \times 5 \times 5$ supercell of UN from a molecular dynamics trajectory at respective temperatures.}
    \label{fig:UNEXAFS}
\end{figure}

As in the previous UO$_2$ case, we use synthetic target spectra for U and N at a single low temperature of 50~K for the refinement procedure. Once again, we verify that the agreement with synthetic spectra also improves agreement with the experimental U EXAFS spectra, even at higher temperatures. In Figure~\ref{fig:UNEXAFS}, we compare the U EXAFS spectra for unrefined and refined MLIPs with respect to experimental EXAFS spectra at (a) 80~K and (b) 300~K. At a high level, the unrefined and refined MLIPs reproduce almost every broad oscillation present in the experimental EXAFS spectra. The EXAFS peak positions of the refined MLIP align slightly better with experimental data compared to the unrefined MLIP, as evidenced by the moderately lower RMSE value. However, the simulated spectra for both the unrefined and refined MLIPs are smooth and do not capture certain sharp features/peaks found in the experimental spectra. We attribute this smoothness to the cutoff radius of 6~\AA~used to calculate the simulated spectra. Increasing the cutoff can capture more nuanced structural information, resulting in more detailed EXAFS spectra~\cite{10.1063/5.0135944, doi:10.1021/acs.inorgchem.3c01346}. However, a greater cutoff radius increases the number of scattering paths from distant coordination shells, which dramatically increases the computational cost of calculating the EXAFS spectra. Selecting a 6~\AA~ cutoff radius is adequate for resolving key structural features in EXAFS while keeping computational costs manageable during the refinement process.

As in the UO$_2$ case, the synthetic target EXAFS spectrum of uranium in UN has a larger amplitude compared to the experimental spectra. The synthetic spectra are calculated for a perfect crystal structure, which eliminates the dampening of peaks due to atomic vibrations. Despite the larger amplitude of target spectra, the refined MLIP actually predicts a slight reduction in peak amplitude (see Figure S17), which improves agreement with the experimental EXAFS spectrum (Figure~\ref{fig:UNEXAFS}). The refinement also causes a slight leftward shift in both U and N EXAFS spectra, mainly in the range of the wavenumber k = 8 to 15. The leftward shift of the EXAFS peaks leads to an increase in the interatomic distances, signifying a decrease in the interatomic forces. With more flexible vibrations, the mean-square-displacement of UN atoms increases, which slightly dampens the spectra. The evidence of weaker interatomic forces is supported by certain properties used to validate the refined MLIP, discussed below.

To validate the refined MLIP for UN against structural properties, Table~\ref{tab:elasticUN} reports the elastic constants at 290~K using the unrefined and refined MLIPs as well as the experimental values at 290~K~\cite{Salleh1986}. The values for $C_{12}$ and $C_{44}$ show an improvement, while the $C_{11}$ value is underestimated after refinement.

\begin{table}[ht]
\centering
\caption{Elastic constants for UN, calculated with unrefined and refined MLIPs at 290~K, compared with the experimental values at 290~K from Salleh et al.~\cite{Salleh1986}.}
\label{tab:elasticUN}
\begin{tabular}{@{}lcccccc@{}}
\toprule
Elastic constants & $C_{11}$ (GPa) & $C_{12}$ (GPa) & $C_{44}$ (GPa) \\ 
\midrule
 Experiment & 423.9 $\pm$ 0.6 & 98.1 $\pm$ 0.9 & 75.7 $\pm$ 0.2 \\
 Unrefined MLIP & 381.59 & 123.30 & 47.65 \\
 Refined MLIP & 351.47 & 116.35 & 51.33 \\
\bottomrule
\end{tabular}
\end{table}

\begin{figure*}
    \centering
    \includegraphics[width=1.0\linewidth]{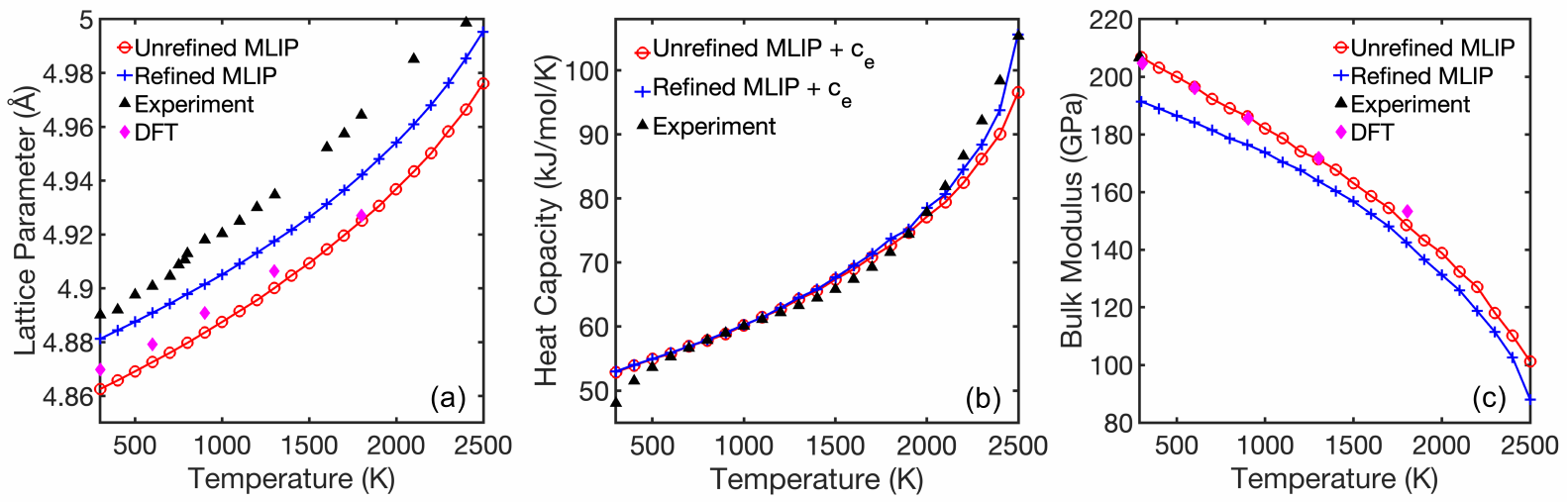}
    \caption{Temperature dependence for thermophysical properties of UN, (a) lattice parameter, (b) heat capacity, and (c) bulk modulus over a range of temperature from 300 K to 2500 K. Predictions obtained using refined and unrefined MLIPs are compared with the experimental values. The experimental lattice parameters and heat capacity are from Hayes et al.~\cite{HAYES1990262}. The experimental bulk modulus data point at 290~K is from Salleh et al.~\cite{Salleh1986}. The heat capacity calculated using the unrefined and refined MLIPs is corrected for the electronic contribution using DFT-computed values for the ferromagnetic UN (c$_e$)~\cite{SZPUNAR2020109636}.}
    \label{fig:UNproperties}
\end{figure*}

The temperature-dependent properties calculated using the unrefined and refined MLIPs for UN in the temperature range between 300~K to 2500~K are shown in Figure~\ref{fig:UNproperties}. Specifically, we calculate: (a) lattice parameter, (b) heat capacity, and (c) bulk modulus. Available experimental values are added for reference. The lattice parameters (Figure~\ref{fig:UNproperties}(a)) calculated using the unrefined MLIP are underestimated relative to the experimental values~\cite{HAYES1990262} across the temperature range. Kocevski et al. also report underestimated lattice parameters of $\approx$0.5\% for UN from DFT-MD simulations~\cite{KOCEVSKI2023154241}, demonstrating that the disagreement between the unrefined MLIP and experiment arises primarily from the DFT training dataset. Refinement reduces the underestimation of lattice parameter with respect to experimental data over the entire temperature range, with a similar degree of improvement to previous work using differentiable trajectory reweighting~\cite{Röcken2024}.

Similar trends for heat capacity are predicted by both the unrefined and refined MLIPs as shown in Figure~\ref{fig:UNproperties}(b). The refined MLIP shows some improvement over the unrefined MLIP at temperatures above 2000~K with respect to experimental data. The bulk modulus as function of temperature calculated with the refined MLIP in Figure~\ref{fig:UNproperties}(c) is lower in comparison to the unrefined MLIP over the entire temperature range. The decrease in bulk modulus is consistent with weaker interatomic forces. However, the refinement process appears to have slightly over-corrected the bulk modulus at low temperatures as the refined MLIP underpredicts the experimental data point. Unfortunately, experimental bulk modulus data are not available at temperatures above 300~K for further validation.

Finally, we calculate point defect energies for UN and report them in Table~\ref{tab:defectUN}. We compare the unrefined and refined MLIP values with DFT. Both MLIPs predict point defect energies that are in agreement with DFT, yet the refined MLIP shows a slight improvement for the uranium and nitrogen Frenkel pair defects compared to the predictions by the unrefined MLIP. This confirms again that despite the being re-trained to shifted atomic energies, the refinement process does not adversely affect the defect energies.

\begin{table}[ht!]
\centering
\caption{Point defect energies for UN predicted with unrefined and refined MLIPs, and compared with the calculated DFT values from Ref. ~\cite{alzatevargas2024machinelearninginteratomicpotential} }
\label{tab:defectUN}
\begin{tabular}{@{}lcccccc@{}}
\toprule
Defect energies (eV) & Schottky defect & Frenkel Pair (U) & Frenkel Pair (N) \\ 
\midrule
 DFT  & 5.15 & 9.46 & 5.04 \\
 Unrefined MLIP & 4.76 & 11.12 & 5.88 \\
 Refined MLIP & 4.58 & 10.71 & 5.58 \\
\bottomrule
\end{tabular}
\end{table}

Certain physical properties, in particular the bulk modulus, are extremely sensitive to the hyperparameters used for the pre-trained MLIP. If the hyperparameters are not chosen judiciously, the refined MLIP may inherit some deficiencies from the unrefined MLIP that are only exacerbated during the refinement process. Since, the refinement process does not retrain to atomic forces, an important hyperparameter is the energy-to-force weight ratio in the loss function for pre-training. For this purpose, we developed another UN MLIP pre-trained with an energy-to-force weight ratio of 1:100 in the loss function, instead of the 1:1 ratio used in previous work~\cite{alzatevargas2024machinelearninginteratomicpotential}, but keeping all other hyperparameters the same. Subsequently, we repeat the refinement process and compare the 1:100 refined MLIP with the 1:1 refined MLIP. Compared to the 1:1 refined MLIP, the 1:100 refined MLIP has a more accurate thermal expansion coefficient across the temperature range and a more stable bulk modulus with respect to refinement generations (see Figure S18). However, certain defect energies and elastic constants are worse for the 1:100 refined MLIP compared to the 1:1 refined MLIP (see Tables S5 and S6). Although these preliminary results suggest that pre-training with a larger force weight can be beneficial for subsequent EXAFS refinement, more research is needed to determine whether this observation applies to other systems.

\section*{Conclusions}

In this work, we used a macroscopic experimental observable--EXAFS spectra--to refine the MLIP of two nuclear fuel materials, UO$_2$ and UN, that were pre-trained using first-principles calculations (DFT+U and DFT, respectively). The choice of improving MLIPs developed for nuclear fuel materials can be useful for accurate atomic-level predictions that can improve mechanistic and multi-physics reactor models used in accelerated fuel qualification--a methodology that aims to reduce the number of nuclear integral tests required in order to qualify a new fuel. Furthermore, this work presents a real-world application typical of computational materials modeling where the DFT+U/DFT calculations are qualitatively reliable for pre-training purposes.

Although the pre-trained MLIPs show good agreement with the respective DFT/DFT+U energies and forces, the agreement with certain experimental properties, e.g., lattice parameter, is limited by the accuracy of the DFT/DFT+U training data. Because EXAFS spectra strongly depend on the interatomic distances, coordination number, and atomic arrangement around an absorbing atom, improving the MLIP to match the target EXAFS spectra particularly improves structural properties. Other properties, such as diffusion coefficient and defect energies remain relatively unchanged, demonstrating that the re-trained MLIP does not ``forget'' the original learning from the DFT/DFT+U dataset. Both, UO$_2$ and UN refined MLIPs are tested for properties such as elastic constants, temperature-dependent lattice parameters, heat capacity, bulk modulus, and defect energies. Additionally, the refined UO$_2$ MLIP is tested for the activation energy of oxygen diffusion and phonon dispersion spectra. 

In general, both MLIPs retain the robustness from pre-training to DFT/DFT+U data while improving the agreement with experimental data. In particular, the predictions for the UO$_2$ refined MLIP improve significantly for the lattice parameter and bulk modulus over the entire temperature range, while marked improvement is also observed for heat capacity at high temperatures, thermal expansion coefficient at high temperatures, elastic constants at $\approx$300~K, and the experimental EXAFS spectra at both 300~K and near the refinement temperature. For UN, the improvement is most pronounced for the lattice parameter over the entire temperature range, while improvement is also observed for heat capacity at high temperatures, elastic constants at $\approx$300~K, and the experimental EXAFS spectra at both 300~K and near the refinement temperature.

The refinement method is based on a re-weighting technique combined with transfer learning. The refinement process is robust, demonstrating systematic improvement of the MLIP and smooth convergence. Although re-training is performed using EXAFS spectra at a single low temperature, the refined MLIP maintains broad applicability, delivering accurate predictions even at significantly higher temperatures.

This work advances the limits of training MLIPs only to quantum mechanical data by refining a pre-trained MLIP to match experimental data. Future work will investigate simultaneous training to both data sources and/or multiple experimental sources. The current refinement method is limited to properties that can be computed for every atom for a given snapshot in a simulation, such as EXAFS and RDF spectra. The procedures pioneered in this paper have a wide range of applicability, including to complex microstructures, e.g., interfaces and grain boundaries, as well as to molecular systems. Future work will investigate the application of this approach to other properties, such as angular distribution functions and pressure-volume equations of state.

\section*{Simulation Details}\label{sec:simulationDetails}

\subsection*{Generating EXAFS Spectra} 

All EXAFS spectra calculations are performed using the $\textit{ab initio}$ real-space multiple-scattering code FEFF8.5~\cite{PhysRevB.58.7565,RevModPhys.72.621,REHR2009548}. The scattering paths only within the radius 6~\AA~of the absorbing atom for both UO$_2$ and UN are considered. To develop the target synthetic EXAFS spectra, FEFF calculations are performed for uranium and oxygen atom in a $4 \times 4 \times 4$ supercell of the UO$_2$ crystal (768 atoms) and, for uranium and nitrogen atoms in a $5 \times 5 \times 5$ UN supercell (1000 atoms). EXAFS corresponding to the K-edge absorption is calculated for O and N, while the L3-edge spectra is calculated for U in both materials. The self-consistent field (SCF) method for calculations improves the agreement of the synthetic EXAFS spectra with the experimental spectra. However, performing SCF is computationally expensive, hence we did not implement SCF for any target spectrum or during the refinement process. SCF was only employed when computing a spectrum for comparison with the experimental EXAFS spectrum. The FEFF parameters used to generate the target synthetic EXAFS spectra using perfect crystal, and the EXAFS spectra of the MD trajectories during the refinement process are kept consistent. FEFF input headers are provided in Supporting Information.

\subsection*{MLIP Refinement} 

The MLIPs for UO$_2$ and UN pre-trained to DFT+U and DFT datasets, respectively, were refined by transfer learning to match the synthetic target EXAFS spectra. Ideally, these target EXAFS spectra would be obtained from experiments. Unfortunately, experimental EXAFS spectra are available only for the uranium atoms in UO$_2$ and UN. For this reason, we generated synthetic target EXAFS spectra for both elements (U and O in UO$_2$, and U and N in UN) using an ideal stoichiometric perfect crystal. To improve the agreement between synthetic spectra and experimental spectra at finite temperature, the perfect crystal is constructed using the experimental lattice parameter at 75~K for UO$_2$ and at 50~K for UN.

MLIP refinement is a three-step process. First, molecular dynamics simulations are performed with the MLIP undergoing refinement, i.e., the original unrefined MLIP for the first refinement generation and preliminary refined MLIPs for subsequent generations. Second, the atomic energy shifts ($\Delta \vec{E}$) are optimized to improve agreement between the reweighted spectra and the target EXAFS spectra (by minimizing the loss function equation~\ref{lossfun}). Third, the MLIP is re-trained using transfer learning to the shifted atomic energies.

MD simulations were performed to generate configurations for EXAFS spectra calculations in the Large-scale Atomic/Molecular Massively Parallel Simulator (LAMMPS)~\cite{THOMPSON2022108171} using a 4 $\times$ 4 $\times$ 4 supercell of UO$_2$ (768 atoms) and 5 $\times$ 5 $\times$ 5 supercell of UN (1000 atoms). First, the atomic positions are relaxed to minimum energy configuration. Then, initial velocities are assigned based on a Maxwell-Boltzmann distribution. Next, isobaric-isothermal (NPT) simulations are performed for 16 ps with a timestep of 0.2~fs. A Nos{\'e}-Hoover barostat maintains a constant pressure of 0~bar with damping at every 1000 timesteps (0.2~ps). A Nos{\'e}-Hoover thermostat maintains constant temperature of 75~K for UO$_2$ and 50~K for UN, with damping at every 100 timesteps (0.02~ps)~\cite{10.1063/1.449071}. After equilibration, eight independent production simulations in the NPT ensemble are performed for 48 ps. Three atomic configurations (snapshots) and the corresponding atomic energies (E$_0$) are saved every 16~ps for each of the eight MD runs. Because each snapshot contains either 768 (UO$_2$) or 1000 (UN) atoms, the total number of atomic energies in the refinement dataset is 18432 for UO$_2$ and 24000 for UN.

EXAFS spectra are calculated for each atom in the refinement dataset using the FEFF8.5 code~\cite{PhysRevB.58.7565,RevModPhys.72.621,REHR2009548}. The spectra are computed only for the wavevector range k = 2.05 to 14.95, with an interval of 0.05. Hence, each spectrum is an array of size 259, consistent with the dimensions of the target synthetic spectra and the experimental spectra. Two loss functions (Equation~\ref{lossfun}) are minimized separately for U atoms and O atoms in UO$_2$ to obtain the shifted atomic energies. 
In Equation~\ref{lossfun}, the size of vector $\vec{S}_{\rm tar}$ is equal to the length of a single spectrum. The number of rows in $\bm{S}_{\rm sim}$ and $\bm{S}_{\rm rw}$ corresponds to the length of a single spectrum, and the number of columns is equal to the total number of spectra. For materials with a single element, the total number of spectra is equal to the total number of atoms in the refinement dataset, and a single loss function is defined. Both systems studied here have two element types, each with a different target EXAFS spectrum, for which two loss functions are defined. Therefore, the number of columns in $\bm{S}_{\rm rw}$ in each loss function represents the number of atoms of the same element type in the refinement dataset. The length of the re-weighting vector ($\vec{w}$) is also equal to the number of same-element atoms in the refinement dataset. The simulated spectra matrix ($\bm{S}_{\rm sim}$) contains data from all selected snapshots and has dimensions = 259 $\times$ 6144 for the loss function corresponding to U, and dimensions = 259 $\times$ 12288 for the loss function corresponding to O in UO$_2$. Similarly, two loss functions are minimized for U atoms and N atoms in UN, where both U and N have ($\bm{S}_{\rm sim}$) dimensions = 259 $\times$ 12000.

Our refinement process utilizes transfer learning to avoid overfitting to the refinement dataset and so that the refined MLIP retains some of the information learned from the DFT/DFT+U training dataset. This approach consists of three key aspects. First, the refined MLIP is initialized with the pre-trained MLIP parameters. Second, retraining is performed with a small learning rate (10$^{-4}$) for a maximum of five total epochs. Third, certain MLIP parameters are not fine-tuned during retraining. Specifically, for UO$_2$, the atomic layers (0 to 3) are kept frozen. For UN, however, no layers were frozen because the pre-trained MLIP required smaller changes and contains nearly five times fewer model parameters than the UO$_2$ MLIP. Thus, the refinement process for UN was sufficiently stable without freezing any layers. 

We tested the efficiency of our refinement process by varying the amount of atomic energies in the refinement dataset by an order of magnitude. Specifically, we tested simulating systems between 324 and 2592 atoms of UO$_2$ and saving between 2 and 20 snapshots per MD simulation. In all cases, the refinement process converges to the experimental lattice parameter at 300~K after approximately six generations of MLIP refinement. However, the larger datasets (more snapshots and/or more atoms) resulted in significantly more overfitting (without freezing any MLIP parameters), as evidenced by extremely poor prediction of lattice parameters at higher temperatures, bulk modulus at low temperatures, and heat capacity over the entire temperature range. For this reason, we recommend simulating relatively small systems and saving only a few snapshots per simulation.

Most of the refinement process is embarrassingly parallel and, thus, scales linearly with respect to the number of MD simulations or the number of saved snapshots per trajectory. Specifically, each MD simulation is performed in parallel on a single GPU, while FEFF calculations are performed in parallel on a single CPU. Thus, the real time to solution for MD and FEFF remains constant when increasing the number of GPUs and CPUs with respect to the size of the refinement dataset. However, our current code does not allow for multi-GPU training of HIP-NN. Fortunately, the retraining step is relatively fast as only a few parameter updates are performed each iteration. Future implementation of multi-GPU training will improve scalability for the MLIP refinement as well, resulting in an overall process with completely linear scaling.

\subsection*{Molecular Dynamics Simulations} 

All MD simulations of temperature-dependent properties reported in this work are performed with the LAMMPS software, unless stated otherwise~\cite{THOMPSON2022108171}. We employed a larger supercell $12 \times 12 \times 12$ for both systems (20736 atoms in UO$_2$, 13824 atoms in UN) to avoid finite-size effects and provide better statistics. A set of MD simulations are performed every 25~K for 20~ps using the isobaric-isothermal ensemble (NPT) with constant pressure (0 bar) and constant temperature in a range between 225~K to~3075 K for UO$_2$ and 225~K and 2575~K for UN. Timestep was set to 2~fs. The Nos{\'e}-Hoover thermostat was used with damping factor of 0.1~ps and barostat with anisotropic pressure control and a damping factor of 0.5~ps~\cite{10.1063/1.449071} for all simulations. The lattice parameters are extracted from simulation by averaging the last 8~ps of simulation. To compute heat capacity $(c_p)$, the enthalpy (H) was obtained by averaging the last 8~ps of each run and used to calculate the partial derivative of enthalpy with respect to temperature at constant pressure as:

\begin{equation}
c_p = \frac{1}{12} \left( \frac{\partial H}{\partial T} \right)_P
\end{equation}
where 12 is the number of repeated conventional cell units. To estimate this partial derivative, the slope of H with respect to T is calculated from the linear fit, using the seven data points between T-75~K and T+75~K.

Bulk modulus is calculated every 100 K between 300~K and 3000~K for UO$_2$, and from 300~K to 2500~K for UN. The nominal volume $(V_0)$ is calculated by averaging the last 4~ps of the NPT MD run at each respective temperature. Subsequently, these equilibrated structures are subject to compression and expansion by 1\% and 2\% each, with respect to the nominal volume $V_0$. The compression and expansion MD simulations are performed using the NVT ensemble (constant volume and temperature). Each NVT simulation is also held for 20~ps using a timestep of 2~fs. The pressure for a given compression or expansion is averaged over the last 4~ps of each run. The bulk modulus $(K)$ is calculated as:

\begin{equation}
K = -V \frac{dP}{dV} \Bigg|_{V = V_0}
\end{equation}
where the partial derivative of pressure with respect to volume is evaluated at the nominal volume $V = V_0$. To estimate this partial derivative, the slope of P with respect to V at a given T is calculated from a linear fit using five data points--two from compression, two from expansion, and one from zero pressure--at the same simulation temperature.

To calculate the near room temperature elastic constants, a 4 $\times$ 4 $\times$ 4 supercell of UO$_2$ (768 atoms), and a 5 $\times$ 5 $\times$ 5 supercell of UN (1000 atoms) is equilibrated in the NPT ensemble at 296~K for UO$_2$ and 290~K for UN for 500~ps, with a timestep of 1~fs. Nos{\'e}-hoover thermostat and barostat was used to held the temperature and pressure constant. The born/matrix method was used to calculate the tensor matrix~\cite{ZHEN2012261}.

The diffusion coefficients $(D)$ for oxygen in UO$_2$ is calculated using the Nernst–Einstein equation:

\begin{equation}
D = \frac{1}{6} \frac{d}{dt} \langle r^2(t) \rangle
\end{equation}
where $\langle r^2(t) \rangle$ is the mean-square-displacement (MSD) and $t$ is time. This derivative is calculated as the linear slope of MSD with respect to time (see Figure S16). To calculate MSD from MD, the UO$_2$ structures were equilibrated in the NPT ensemble for 20~ps from 2100~K to 3000~K with 100~K increments. 
We then perform a constant energy simulation using the NVE ensemble for 1~ns with a timestep of 2~fs. Transport properties are best computed with the NVE ensemble so that the dynamics are not impacted by the barostat or thermostat~\cite{Maginn_Messerly_Carlson_Roe_Elliot_2018}. MSD values are recorded every 200~fs from the generated trajectory. 

The activation energy $E_a$ is obtained from a linear fit of the diffusion coefficient as a function of temperature following an Arrhenius equation: 

\begin{equation}
\ln(D) = \ln(D_0) - \frac{E_a}{k_{\rm B}} \frac{1}{T}
\end{equation}
where $D$ is the diffusion coefficient, $D_0$ is a pre-exponential factor, $T$ is the temperature, and $k_{\rm B}$ is the Boltzmann constant. 

Point defect energies are evaluated for a 4 $\times$ 4 $\times$ 4 UO$_2$ supercell (768 atoms) and a 5 $\times$ 5 $\times$ 5 UN supercell (1000 atoms). For the Schottky defect (SD) in UO$_2$, a cation U$^{4+}$ and two anions O$^{2-}$ are removed from the crystal structure. In the case of UN, a cation U$^{3+}$ and an anion N$^{3-}$ are removed from the crystal structure. The SD defect energy $(E_{\rm SD})$ is calculated as: 

\begin{equation}
E_{\text{SD}} = E_{\text{defected}} - \left(\frac{n_{\text{defected}}}{n_{\text{perfect}}}\right) E_{\text{perfect}}
\end{equation}
where $n_{\text{defected}}$ and $n_{\text{perfect}}$ is the number of atoms in the defected and perfect system, respectively, and $E_{\text{defected}}$ and $E_{\text{perfect}}$ is the energy of the defected and perfect crystal structure, respectively. 

Frenkel pair defect energies (FP) are defined separately for a cation and an anion. In either case, a charged species is removed from its lattice position and placed at an interstitial site, thus retaining a charge-neutral system. For both UO$_2$ and UN, the interstitial position to place the cation/anion is selected randomly. Since $n_{\text{defected}}$ is equal to $n_{\text{perfect}}$ in the case of FP defect, the FP defect energy $(E_{\rm FP})$  is calculated as:

\begin{equation}
E_{\text{FP}} = E_{\text{defected}} - E_{\text{perfect}}
\end{equation}

The UO$_2$ phonon dispersion band at 0~K is computed for a $10 \times 10 \times 10$ supercell using the finite difference method as implemented in the Atomic Simulation Environment (ASE) package~\cite{HjorthLarsen_2017}.

\section*{Appendix}

\subsection*{Aluminum Example}
Our refinement process to experimental data is not limited to only EXAFS data, but can be applied to any experimental observable where the value is an average of responses from individual atoms. To highlight this fact, early in the development of the methodology presented in this manuscript, we implemented our refinement approach to compare with the refinement procedure of Matin et. al.~\cite{doi:10.1021/acs.jctc.3c01051} Specifically, we refined a DFT-level pre-trained MLIP for aluminum using experimental liquid RDF spectra. Similar to Matin et al., our refined MLIP predicts improved RDF spectra as well as improved diffusion coefficients. However, our refinement process based on modified reweighting has two key advantages over the procedure in Matin et al. First, the MLIP parameters are modified directly themselves, rather than introducing an additional pairwise potential. Second, our methodology is capable of refining to multiple temperatures simultaneously. In this section, we present results for refining the aluminum MLIP.

The unrefined MLIP is initially trained to an existing DFT-level aluminum dataset.\cite{aniAl} For this example, the MLIP is also based on the HIP-NN architecture with similar hyperparameters to the UO$_2$ and UN MLIPs (see Supporting Information). Next, MD simulations are performed at two different temperatures for which the experimental RDFs are available, namely 1023~K and 1323~K.\cite{Waseda1980} The simulations were equilibrated for 200~ps using a 0.5~fs timestep. After equilibration, production MD simulations are run for 250~ps in the isobaric-isothermal (NPT) ensemble. During production, a binned RDF for every atom is saved every 5~ps. These binned atomic RDFs (ARDFs) extend to 10.0\AA~and contain 200 evenly spaced bins. The binned ARDFs are simply a count of the number of atoms around the central atom in a given radius $(r)$, i.e., between r and r+$\Delta$ r. To improve statistical averaging and independent sampling, four MD simulations were run at each temperature. Each simulation contained 2048 atoms, resulting in a total of 819,200 binned atomic RDFs. ARDFs computed for a single atom from a single snapshot are intrinsically noisy. However, the RDF remains smooth and can be recovered by averaging over all snapshots and all atoms.
\begin{equation}
\label{eq:atomicRDF}
RDF(r) = \frac{1}{N}\sum_{i=1}^{N} \frac{ARDF_i(r)}{2\rho_i \pi r^2\Delta r}
\end{equation}
where $N$ is the number of binned ARDFs, $r$ is the distance at which the ARDF is evaluated, and $\rho_i$ is the instantaneous number density of the simulation at the corresponding snapshot.

Utilizing this reconstruction rule, the individual ARDFs were employed in Eq. \ref{eq:Srw8} as $\bm{S}_{\rm sim}$ and the same reweighting procedure was employed to determine the $\Delta \vec{E}$ values. This procedure is performed in parallel at each temperature independently. The $\Delta \vec{E}$ values are then added to the HIP-NN atomic energies to obtain the shifted atomic energies for each atomic configuration from the MD simulations, exactly as described in the main text. HIP-NN was then trained to the shifted atomic energies for 5 epochs with a learning rate of 1e-4. Once the next generation of MLIP was trained, new MD simulations were performed and the whole refinement process was repeated a total of nine times.

\begin{figure}
    \centering
    \includegraphics[width=1.0\linewidth]{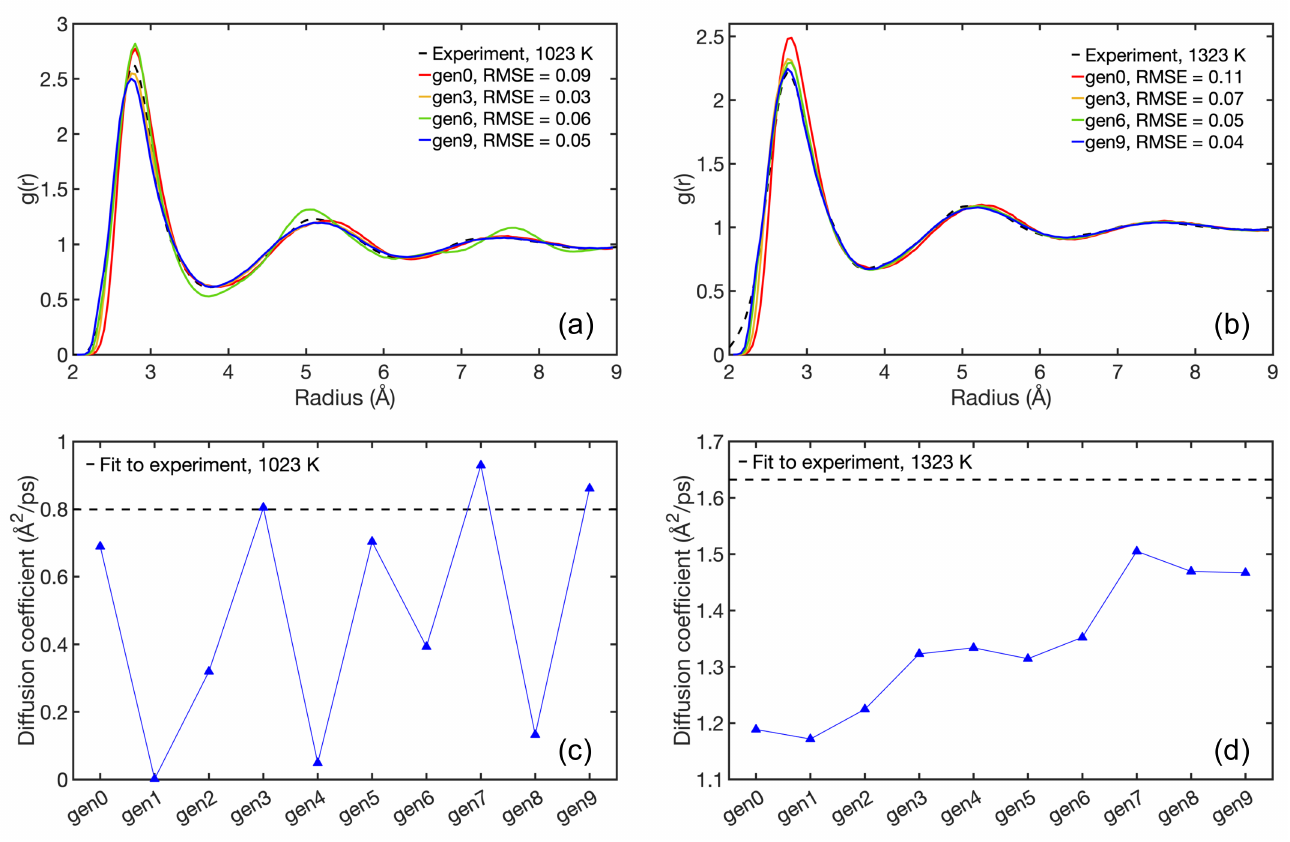}
    \caption{Comparison of RDF spectra (top) and diffusion coefficients (bottom) at 1023~K (left) and 1323~K (right) as measured by experiment and simulated with MLIP at each refinement iteration. Here, gen0 corresponds to the unrefined MLIP pre-trained to DFT, gen9 corresponds to the final RDF-refined MLIP, and genx with $0<$x$<9$ corresponds to all the other generations during the refinement process. The simulated RDF spectra are obtained by averaging over all 2048 Al atoms in four independent MD trajectories at the respective temperatures. The experimental RDF data come from Waseda~\cite{Waseda1980}. The experimental diffusion coefficients are obtained from an Arrhenius fit to the experimental data of Kargl et al.\cite{Kargl_book_2013} covering a temperature range of 944~K to 1194~K. The uncertainty in the fit to experiment for the diffusion coefficient is much larger at 1323~K than at 1023~K due to the required extrapolation of over 100~K from the experimental temperature range.}
    \label{fig:AX_AL}
\end{figure}

As shown in Figure~\ref{fig:AX_AL}, the refined MLIP predicts a total RDF in good agreement with the experimental data. Further, the simulated diffusion rates show significant improvement to the experimental values. The intermediate MLIPs show systematic improvement in the RDF and diffusion coefficient at the higher temperature with respect to generations of refinement. However, the non-monotonic improvement in the RDF and diffusion coefficient at the lower temperature is due to the refined MLIP occasionally raising the melting point near or above 1023~K, causing a precipitous drop in the diffusivity. However, this increase in the MLIP melting point is immediately corrected with the next generation of experimental refinement. Overall, this further example shows that the refinement procedure is generalizable to different experimental observables and can be applied to multiple experimental measurements (e.g., temperatures) at once.

\section*{Declarations}

\subsection*{Acknowledgments}

The authors appreciate the helpful scientific discussions with Nicholas Lubbers, Sakib Matin, Roxanne Tutchton, Michael Cooper, and Kashi Subedi. S.G. is grateful for the mentorship provided by Oleg Prezhdo and Sergei Tretiak. S.G., L.A.V., B.N., A.v.V., T.G., and R.A.M. acknowledge support from Los Alamos National Laboratory (LANL), an affirmative action/equal opportunity employer, operated by Triad National Security LLC, for the National Nuclear Security Administration of the U.S. Department of Energy under Contract No. 89233218CNA000001. R.A.M. acknowledges the Oak Ridge Leadership Computing Facility at the Oak Ridge National Laboratory, which is supported by the Office of Science of the U.S. Department of Energy under Contract No. DE-AC05-00OR22725.

\subsection*{Funding}

The work at LANL was supported by the LANL Laboratory Directed Research and Development (LDRD) Project 20220053DR. S.G. gratefully acknowledges the resources of the Los Alamos National Laboratory (LANL) Computational Science summer student program. This research used resources provided by the Darwin testbed at LANL which is funded by the Computational Systems and Software Environments subprogram of LANL's Advanced Simulation and Computing program (NNSA/DOE). This research used resources provided by the Los Alamos National Laboratory Institutional Computing Program, which is supported by the U.S. Department of Energy National Nuclear Security Administration under Contract No. 89233218CNA000001.

\subsection*{Author Contributions} 

R.A.M. conceived and supervised the project. S.G. trained, refined and validated MLIP, and analyzed the results. B.T.N. designed the refinement method and wrote the original refinement code. S.K. introduced and implemented the freezing-layer strategy for transfer learning in the HIP-NN code. S.G. and R.A.M. revised the refinement method and updated the refinement code. L.A.V. supervised the training and molecular dynamics calculations. A.v.V. supervised the EXAFS calculations and provided the experimental EXAFS data. T.G. procured funding and oversaw the project. S.G. and R.A.M wrote the main manuscript draft. L.A.V., B.T.N and S.K. edited and reviewed the manuscript.

\subsection*{Competing interests}

The authors declare no competing financial interests.

\bibliographystyle{ieeetr} 
\bibliography{reference.bib}

\newpage
\setcounter{figure}{0}
\setcounter{table}{0}
\renewcommand{\arraystretch}{1.0}
\renewcommand{\thefigure}{S\arabic{figure}}
\renewcommand{\thetable}{S\arabic{table}}
\renewcommand{\thepage}{S\arabic{page}}

\begin{center}
  {\Large\bfseries
   Supporting Information for:\\[0.5\baselineskip]
   Going beyond density functional theory accuracy:\\
   Leveraging experimental data to refine pre-trained machine learning interatomic potentials
  }
\end{center}

\subsection*{Hierarchically Interacting Particle Neural Network (HIP-NN)}
HIP-NN training works by processing the feature vectors for each atom through interaction blocks consisting of interaction layers followed by a number of atomic layers. At $0^{th}$ input layer, the feature vector is simply the one-hot encoding representing the species of an atom $i$ (Eg., U, O, N). At subsequent layers, feature vector ($z_{i,a}$) of atom $i$ has length $a = n_{features}$. At interaction layers, the atom learns from its surrounding atoms within a pre-defined cutoff radius, i.e., the feature vector on one atom mixes with the feature vectors of its neighboring atoms. Mathematically, the input feature vector $z$ is transformed to $z'$ using:
\begin{equation}
{z'}_{i,a}=f\left(\mathcal{I}_{i,a}(z,\mathbf{r})+\sum_{b}W_{ab}z_{i,b}+B_{a}\right).\label{eq:hipnn-layer}
\end{equation}
where $W$ is trainable weight matrix, $B$ is the offset vector and $f$ is the activation function for non-linear transformation. The interaction term $\mathcal{I}_{i,a}(z,\mathbf{r})$ enables the central atom $i$ to receive information from the neighboring atom $j$ within the cutoff radius. In this work, we use HIP-NN with tensor sensitivities\cite{10.1063/5.0142127}, in which the message from $j$ to $i$ are generalized to rank-$l$ tensors, according to:
\begin{equation}
\boldsymbol{\mathcal{E}}_{i,a}^{(\ell)}=\sum_{j}\mathbf{M}_{ij,a}^{(\ell)}=\sum_{j}\mathbf{T}^{(\ell)}(\hat{\mathbf{r}}_{ij})m_{ij,a},\label{eq:env_tensor}
\end{equation}
where $m_{ij,a}$ is interpreted as a scalar message passed from atom $j$ to $i$ that does not contain any angular information. The rank-$\ell$ irreducible Cartesian tensors $\mathbf{T}^{(\ell)}(\mathbf{r})$ can be used to describe functions on the sphere, where the first four tensors are:
\begin{align}
T^{(0)}(\mathbf{r}) & =1\label{eq:monopole}\\
T_{\alpha}^{(1)}(\mathbf{r}) & =r_{\alpha} \\
T_{\alpha\beta}^{(2)}(\mathbf{r}) & =r_{\alpha}r_{\beta}-\frac{1}{3} \delta_{\alpha \beta} r^{2} \\
T_{\alpha\beta\gamma}^{(3)}(\mathbf{r}) & =r_{\alpha}r_{\beta}r_{\gamma}-\frac{1}{5}(\delta_{\alpha\beta}r_{\gamma}+\delta_{\alpha\gamma}r_{\beta}+\delta_{\beta\gamma}r_{\alpha})r^{2}\label{eq:octupole}
\end{align}
Here, $\alpha$, $\beta$, and $\gamma$ are Cartesian indices that take on values of $x$, $y$, and $z$ in some coordinate system, $\delta$ is the Kronecker delta, and $r^2$ is the magnitude squared of the input vector $\mathbf{r}$. The interaction term is a weighted sum of rank-$l$ tensors:
\begin{equation}
\mathcal{I}_{i,a}=\boldsymbol{\mathcal{E}}^{(0)}_{i,a}+t^{(1)}_a|\boldsymbol{\mathcal{E}}_{i,a}^{(1)}|+t^{(2)}_a|\boldsymbol{\mathcal{E}}_{i,a}^{(2)}|+\dots\label{eq:hip-vec-1}
\end{equation}

Unlike interaction layers, the atomic layer locally transforms features on an individual atom. The transformation from the input feature vector $z_i$ to $z_i'$ is given by:
\begin{equation}
\label{atomic_layer}
    z'_{i,a} = f\left(\sum_n w_{ab}z_{i,n} + b_a\right)
\end{equation}
where $w$ is a trainable weight matrix, $b$ is the offset vector, and $f$ is a non-linear activation function.

The hierarchical energy for each atom is obtained from the linear regression of the feature vector prior to each interaction block according to:
\begin{equation}
\label{linear_output_layer}
    E^n_i = \sum_a w_a z_{i,a}+b
\end{equation}
where $w_a$ and $b$ are learnable weights and biases. HIP-NN predicts atomic energy as a sum of $n$ hierarchical energy terms:
\begin{equation}
    E_i = \sum_{n=0}^{n_{\text{int}}-1} E^n_i
\end{equation}
where $E_i^n$ is the $n^{\rm th}$ hierarchical energy for atom $i$, and $n_{\text{int}}$ is the number of interaction blocks. Total energy is simply the sum of atomic energy in an atomic configuration:
\begin{equation}
    E = \sum_{i=1}^{n_{\text{atom}}} E_i
\end{equation}

\subsection*{Loss Function}
Loss function for HIP-NN training is given by:
\begin{align*}
\mathcal{L} = & \ c_1 \times (\text{RMSE}_{\text{energy-per-atom}} + \text{MAE}_{\text{energy-per-atom}}) \\
& + c_2 \times ( \text{RMSE}_{\text{forces}} + \text{MAE}_{\text{forces}}) \\
& + c_3 \times \mathcal{L}2 +  c_4 \times \mathcal{L}_{R}
\end{align*}
where RMSE stands for root-mean-square error and MAE stands for mean absolute error. $\mathcal{L}2$ regularization term is for the weight tensors and $ \mathcal{L}_{R}$ regularization term encourages hierarchical energy terms. $c_1$, $c_2$ are the pre-factors for energy and forces errors, and $c_3$, $c_4$ are the pre-factors for $\mathcal{L}2$ regularization, and $\mathcal{L}_{R}$ regularization, respectively. 

\subsection*{Hyperparameter tuning for UO$_2$ MLIP}

For training purposes, the total dataset is divided into 80$\%$ training data, 10$\%$ testing data, and 10$\%$ validation data.

\begin{table}[ht]
\centering
\caption{To determine the optimal number of features ($n_{\rm features}$) and interaction layers ($n_{\rm int}$) to train UO$_2$ MLIP with the HIP-NN architecture, energy and force RMSE values are computed for the training, testing and validation dataset and compared. Other hyperparameters are kept constant as: soft min = 1.5, soft max = 5, hard max = 5.7, atomic layers = 4, sensitivities = 20, $\ell_{\text{max}}$=1, $c_1$=1, $c_2$=1, $c_3$=0.000001, $c_4$=0.01.}

\begin{tabular}{@{}lcccccc@{}}
    \centering
    \begin{tabular}
    {p{1.2cm}p{1.2cm}p{2cm}p{2cm}p{2cm}p{2cm}p{2cm}p{2cm}}
        \hline
        \multirow{2}{*}{$n_{\rm int}$} & \multirow{2}{*}{$n_{\rm features}$} & 
        \multicolumn{3}{c|}{RMSE energy-per-atom (kcal/mol)}  &
        \multicolumn{3}{c|}{RMSE forces (kcal/mol/Å)} \\
        \cline{3-5} \cline{6-8} 
        & & train & valid & test & train & valid & test \\
        \hline
        1 & 20  & 
        0.075$\pm$0.006 & 
        0.074$\pm$0.006 & 
        0.075$\pm$0.006 & 
        3.636$\pm$0.061 & 
        3.728$\pm$0.187 & 
        3.773$\pm$0.178 \\
        1 & 60 & 
        0.071$\pm$0.016 & 
        0.071$\pm$0.016 & 
        0.073$\pm$0.018 & 
        3.412$\pm$0.225 & 
        3.653$\pm$0.457 & 
        3.660$\pm$0.322 \\
        1 & 120  & 
        0.069$\pm$0.012 & 
        0.069$\pm$0.012 & 
        0.071$\pm$0.012 & 
        3.296$\pm$0.156 & 
        3.654$\pm$0.613 & 
        3.608$\pm$0.319 \\
        2 & 20 & 
        0.067$\pm$0.029 & 
        0.068$\pm$0.028 & 
        0.068$\pm$0.030 & 
        3.024$\pm$0.121 & 
        3.218$\pm$0.394 & 
        3.125$\pm$0.114 \\
        2 & 60  & 
        0.048$\pm$0.004 & 
        0.052$\pm$0.003 & 
        0.052$\pm$0.003 & 
        2.530$\pm$0.046 & 
        2.932$\pm$0.164 & 
        2.904$\pm$0.084 \\
        2 & 120 & 
        0.049$\pm$0.004 & 
        0.053$\pm$0.003 & 
        0.053$\pm$0.004 & 
        2.362$\pm$0.086 & 
        3.016$\pm$0.291 & 
        2.951$\pm$0.093 \\
        \hline
    \end{tabular}
    \label{tab:features_int}
\end{tabular}
\end{table}

When UO$_2$ MLIP is trained to the HIP-NN architecture with 1 interaction layer, the RMSE for energy-per-atom and forces decreases for all 3 datasets (train, valid, test) as the number of features increases (Table~\ref{tab:features_int}). Interestingly, this is not the case when the number of interaction layers is increased to 2. In this case, while the training dataset RMSE decreases with increasing features, the testing and validation dataset RMSE values (for both energy and forces) are minimum for 60 features instead of 120. The complex neural network architecture with 2 interaction layers and 120 features clearly suffers from overfitting. From this analysis, two combinations of $n_{int}$ and $n_{features}$ hyperparameters are selected (interaction layers = 1, number of features = 120) and (interaction layers = 2, number of features = 60). Further, the soft max distance is determined for both sets of hyperparameters (Table~\ref{tab:smax_hmax} and ~\ref{tab:int1feat120}). 

\begin{table}[H]
\centering
\caption{For set 1 \textbf{(interaction layers = 2, number of features = 60)}. To determine the soft max and hard max to train UO$_2$ MLIP with the HIP-NN architecture, energy and force RMSE values are computed for the training, testing and validation dataset and compared. Other hyperparameters are kept constant as: soft min = 1.5, interaction layers = 2, number of features = 60, atomic layers = 4, sensitivities = 20, $\ell_{\text{max}}$=1, $c_1$=1, $c_2$=1, $c_3$=0.000001, $c_4$=0.01.}
\begin{tabular}{@{}lcccccc@{}}
    \centering
    \begin{tabular}
    {p{1.2cm}p{1.2cm}p{2cm}p{2cm}p{2cm}p{2cm}p{2cm}p{2cm}}
        \hline
        \multirow{2}{*}{soft max} & \multirow{2}{*}{hard max} & 
        \multicolumn{3}{c|}{RMSE energy-per-atom (kcal/mol)}  &
        \multicolumn{3}{c|}{RMSE forces (kcal/mol/Å)} \\
        \cline{3-5} \cline{6-8} 
        & & train & valid & test & train & valid & test \\
        \hline
        4 & 4.7  & 
        0.059$\pm$0.030 & 
        0.061$\pm$0.028 & 
        0.062$\pm$0.029 & 
        2.641$\pm$0.178 & 
        2.972$\pm$0.259 & 
        2.974$\pm$0.128 \\
        5 & 5.7 & 
        0.048$\pm$0.004 & 
        0.052$\pm$0.003 & 
        0.052$\pm$0.003 & 
        2.529$\pm$0.046 & 
        2.932$\pm$0.164 & 
        2.904$\pm$0.084 \\
        \hline
    \end{tabular}
    \label{tab:smax_hmax}
\end{tabular}
\end{table}

\begin{table}[ht]
\centering
\caption{For set 2 \textbf{(interaction layers = 1, number of features = 120)}. To determine the soft max and hard max to train UO$_2$ MLIP with the HIP-NN architecture, energy and force RMSE values are computed for the training, testing and validation dataset and compared. Other hyperparameters are kept constant as: soft min = 1.5, , atomic layers = 4, sensitivities = 20, $\ell_{\text{max}}$=1, $c_1$=1, $c_2$=1, $c_3$=0.000001, $c_4$=0.01.}
\begin{tabular}{@{}lcccccc@{}}
    \centering
    \begin{tabular}
    {p{1.2cm}p{1.2cm}p{2cm}p{2cm}p{2cm}p{2cm}p{2cm}p{2cm}}
        \hline
        \multirow{2}{*}{soft max} & \multirow{2}{*}{hard max} & 
        \multicolumn{3}{c|}{RMSE energy-per-atom (kcal/mol)}  &
        \multicolumn{3}{c|}{RMSE forces (kcal/mol/Å)} \\ 
        \cline{3-5} \cline{6-8} 
        & & train & valid & test & train & valid & test \\  
        \hline
        4 & 4.7  & 
        0.068$\pm$0.008 & 
        0.068$\pm$0.009 & 
        0.070$\pm$0.008 & 
        3.334$\pm$0.076 & 
        3.584$\pm$0.324 & 
        3.543$\pm$0.201 \\
        5 & 5.7 & 
        0.069$\pm$0.012 & 
        0.070$\pm$0.012 & 
        0.071$\pm$0.012 & 
        3.296$\pm$0.156 & 
        3.653$\pm$0.613 & 
        3.608$\pm$0.319 \\
        \hline
    \end{tabular}
    \label{tab:int1feat120}
\end{tabular}
\end{table}

The RMSE values for energy-per-atom and forces obtained with set 2 (interaction layers = 1, number of features = 120) are only slightly larger than RMSE values obtained with hyperparameters in set 1 (interaction layers = 2, number of features = 60). The set 1 has relatively much higher computational cost, with only small benefits on accuracy. We balance the accuracy and efficiency, and chose to train the MLIP with hyperparameters in set 2. For set 2, the softmax of 4 offers slightly smaller RMSE values than the softmax of 5, while being efficient at the same time. Hence, set 2 with softmax of 4 and hard max 4.7 is chosen for training the MLIP. 

Further we determine the pre-factors of loss function $c_1$ and $c_2$ to train UO$_2$ MLIP with the HIP-NN architecture, by keeping the chosen hyperparameters fixed, (i.e., interaction layers = 1, features = 120, soft max = 4, hard max = 4.7, and soft min = 1.5, atomic layers = 4, sensitivities = 20, $\ell_{\text{max}}$=1) in table~\ref{tab:hyperparam_c1_c2} below.

\begin{table}[H]
\centering
\caption{RMSE values for energy-per-atom and forces calculated with various combinations of $c_1$ and $c_2$ pre-factors. Other pre-factors are kept constant as: $c_3$=0.000001, and $c_4$=0.01.}
\begin{tabular}{@{}lcccccc@{}}
    \centering
    \begin{tabular}
    {p{1.2cm}p{1.2cm}p{2cm}p{2cm}p{2cm}p{2cm}p{2cm}p{2cm}}
        \hline
        \multirow{2}{*}{$c_1$} & \multirow{2}{*}{$c_2$} & 
        \multicolumn{3}{c|}{RMSE energy-per-atom (kcal/mol)}  &
        \multicolumn{3}{c|}{RMSE forces (kcal/mol/Å)} \\ 
        \cline{3-5} \cline{6-8} 
        & & train & valid & test & train & valid & test \\  
        \hline
        1 & 100  & 
        0.221$\pm$0.122 & 
        0.216$\pm$0.123 & 
        0.219$\pm$0.127 & 
        9.602$\pm$0.066 & 
        9.912$\pm$0.288 & 
        9.873$\pm$0.118 \\
        1 & 10 & 
        0.130$\pm$0.006 & 
        0.130$\pm$0.003 & 
        0.127$\pm$0.007 & 
        9.592$\pm$0.064 & 
        9.928$\pm$0.314 & 
        9.880$\pm$0.114 \\
        1 & 1 & 
        0.124$\pm$0.002 & 
        0.123$\pm$0.006 & 
        0.121$\pm$0.005 & 
        9.651$\pm$0.045 & 
        9.972$\pm$0.307 & 
        9.916$\pm$0.109 \\
        10 & 1 & 
        0.116$\pm$0.003 & 
        0.125$\pm$0.006 & 
        0.123$\pm$0.006 & 
        9.918$\pm$0.060 & 
        10.276$\pm$0.487 & 
        10.249$\pm$0.250 \\
        100 & 1 & 
        0.107$\pm$0.005 & 
        0.141$\pm$0.005 & 
        0.139$\pm$0.006 & 
        11.255$\pm$0.102 & 
        11.505$\pm$0.241 & 
        11.654$\pm$0.311 \\
        \hline
    \end{tabular}
    \label{tab:hyperparam_c1_c2}
\end{tabular}
\end{table}

\begin{figure}[H]
    \centering
    \includegraphics[width=0.8\linewidth]{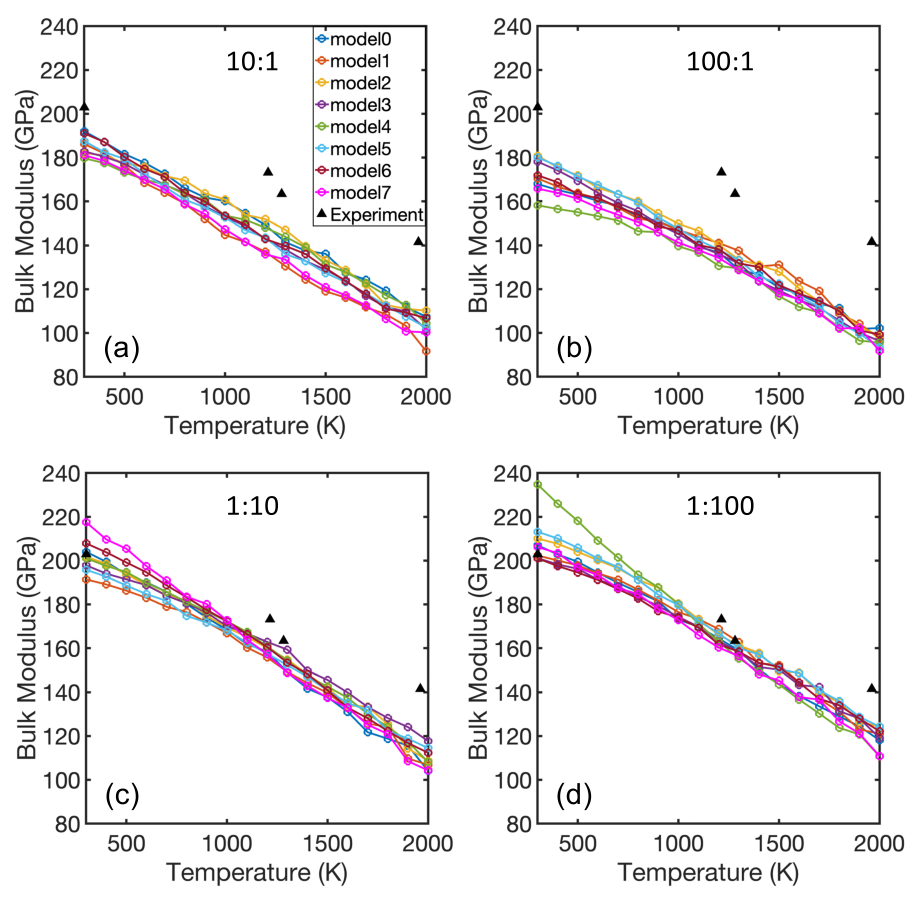}
    \caption{Before pre-training the UO$_2$ MLIP to DFT+U data, different pre-factors for energy and forces root-mean-square error (RMSE) terms in the loss function are evaluated by comparing the bulk modulus calculated using each pre-factor with the experimental values. For each combination of energy-to-force pre-factors, (a) 10:1, (b) 100:1, (c) 1:10, and (d) 1:100, an ensemble of 8 MLIPs is trained, and bulk modulus is calculated for these MLIPs. The bulk modulus calculated with MLIP in (d)1:100 energy-to-force ratio is closest to the experimental results, and hence it is chosen for pre-training MLIP for further refinement.}
    \label{fig:UO2_bm_hyperparam}
\end{figure}

From the RMSE values in Table~\ref{tab:hyperparam_c1_c2}, we find that the MLIP trained with the loss function that has higher weight of the force pre-factor results in smaller RMSE values for forces, and MLIP trained with loss function that has higher weight for the energy pre-factor results in smaller RMSE values for energy. Hence, we consider calculating a temperature dependent physical property such as bulk modulus and compare to experiments to determine the best fit for pre-factors. From Figure~\ref{fig:UO2_bm_hyperparam}, we identify that $c_1$ to $c_2$ ratio of 1:100 has bulk modulus closer to experimental values. Moreover, during the re-training, we are fine-tuning the MLIP parameters with only the atomic energies in the refinement dataset, compared to the unrefined MLIP which is pre-trained to both total energy and atomic forces from the DFT dataset. Hence, we considered training the MLIP with relatively higher pre-factors on forces than energy, i.e., $c_1$=1 and $c_2$=100.

\clearpage
\newpage

\clearpage
\newpage

\subsection*{Hyperparameters for training the UO$_2$ MLIP}
\begin{itemize}
    \item number of features: 120
    \item sensitivities: 20
    \item distance, soft min: 1.5
    \item distance, soft max: 4
    \item distance, hard max: 4.7
    \item number of interaction layers: 1
    \item number of atomic layers: 4
    \item sensitivity type: inverse
    \item cusp regularization: 0.01
    \item $c_1$ = $1$, $c_2$ = $100$, $c_3$ = $10^{-6}$, $c_4$ = $10^{-2}$
    \item learning rate: 0.001
    \item tensor rank, $\ell_{\text{max}}$: 1
    \item total training parameters: 136456 
\end{itemize}

\subsection*{Hyperparameters for training the UN MLIP}
\begin{itemize}
    \item number of features: 60
    \item sensitivities: 20
    \item distance, soft min: 1.5
    \item distance, soft max: 4
    \item distance, hard max: 5
    \item number of interaction layers: 1
    \item number of atomic layers: 3
    \item sensitivity type: inverse
    \item cusp regularization: 0.001
    \item $c_1$ = $1$, $c_2$ = $1$, $c_3$ = $10^{-6}$, $c_4$ = $10^{-1}$. 
    \item learning rate: 0.001
    \item tensor rank, $\ell_{\text{max}}$: 1
    \item total training parameters: 28576 
\end{itemize}

\subsection*{Hyperparameters for training the Aluminum MLIP}
\begin{itemize}
    \item number of features: 100
    \item sensitivities: 20
    \item distance, soft min: 1.25
    \item distance, soft max: 7
    \item distance, hard max: 7.5
    \item number of interaction layers: 1
    \item number of atomic layers: 3
    \item sensitivity type: inverse
    \item cusp regularization: 0.001
    \item $c_1$ = $1$, $c_2$ = $100$, $c_3$ = $10^{-6}$, $c_4$ = $1$
    \item learning rate: 0.001
    \item tensor rank, $\ell_{\text{max}}$: 2
    \item total training parameters: 73356
\end{itemize}
\clearpage
\newpage
\subsection*{FEFF input headers to generate target EXAFS spectra}

\subsubsection*{For Uranium atoms in both UO$_2$ and UN}
\begin{itemize}
  \renewcommand\labelitemi{}
  \item HOLE      4   1 
  \item CONTROL   1         1     1     1 
  \item PRINT     1      0     0     0
  \item EXCHANGE 0 0.0 0.0 -1 
  \item RPATH 6.0 
  \item EXAFS 20 
\end{itemize}

\subsubsection*{For Oxygen atoms in UO$_2$}
\begin{itemize}
  \renewcommand\labelitemi{}
  \item HOLE      1   1  
  \item CONTROL   1         1     1     1
  \item PRINT     1      0     0     0
  \item EXCHANGE 0 0.0 0.0 -1 
  \item RPATH 6.0 
  \item EXAFS 20 
\end{itemize}

\subsubsection*{For Nitrogen atoms in UN}
\begin{itemize}
  \renewcommand\labelitemi{}
  \item HOLE      1   1
  \item CONTROL   1         1     1     1
  \item PRINT     1      0     0     0
  \item EXCHANGE 0 0.0 0.0 -1
  \item RPATH 6.0
  \item EXAFS 20 
\end{itemize}

The above FEFF header is followed by the potentials and atomic positions section in the FEFF input file. The FEFF header was kept same for every MLIP generation during the entire refinement process. However, to compare with experimental EXAFS spectra (in Figure 2 and 6 in main paper), an expensive and more accurate calculation with self-consistent field (SCF) was done to generate U EXAFS spectra for unrefined and refined MLIPs. To include SCF, an additional header (SCF 6.0 0 30 0.2 1) was added to the FEFF input script. 
\clearpage
\newpage

\begin{figure}[H]
    \centering
    \includegraphics[width=0.6\linewidth]{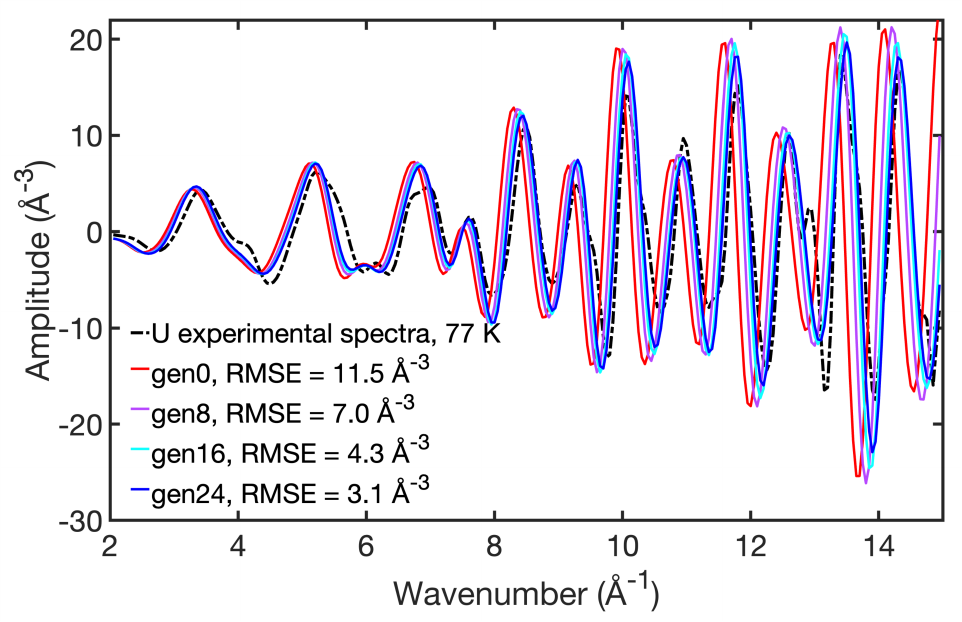}
    \caption{Preliminary re-training attempts to match the experimental EXAFS spectra only for U in UO$_2$ at 77 K. In the absence of experimental EXAFS spectra for O, the atomic energy shifts corresponding to oxygen atoms are assumed to be zero. Expensive SCF method in FEFF is used to generate the EXAFS spectra at each generation. Unrefined MLIP had 1:1 energy-to-force weight ratio. Cusp regularization factor for training in HIP-NN was not included.}
    \label{fig:onlyU_re-training_77K}
\end{figure}

\begin{figure}[H]
    \centering
    \includegraphics[width=1.0\linewidth]{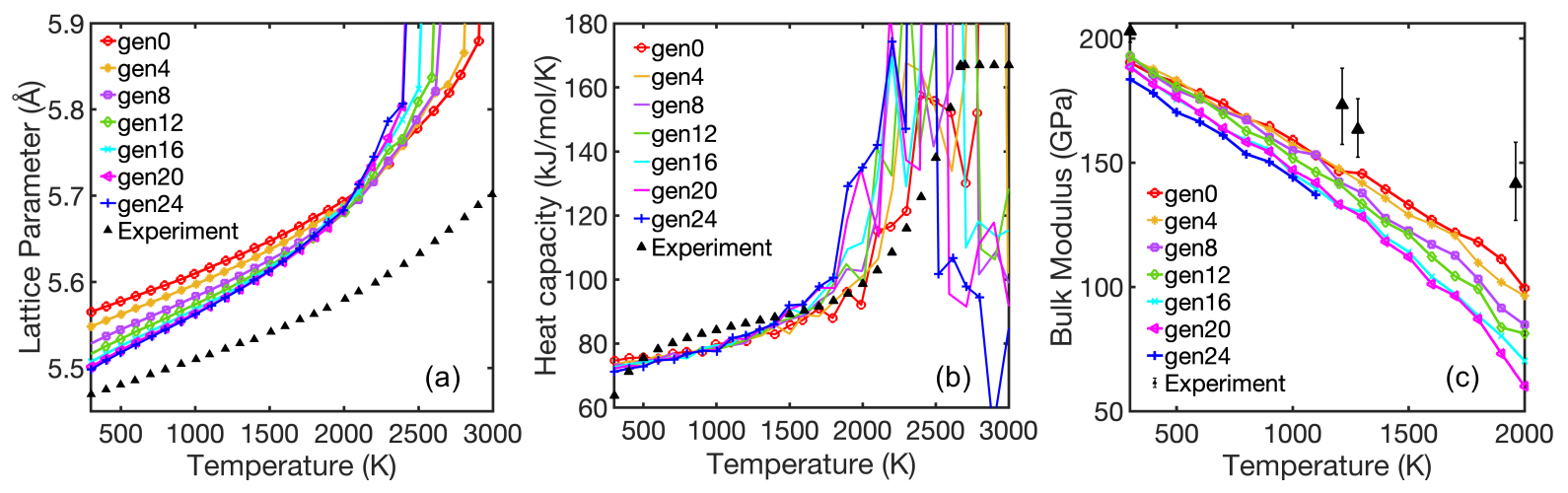}
    \caption{Thermophysical properties obtained with the unrefined MLIP (gen0), refined MLIP (gen24) and in-between generation MLIPs (gen$x$, where $0<x<24$), when re-training is done to match the experimental EXAFS spectra only for U in UO$_2$ at 77 K. The unrefined MLIP (gen0) is pre-trained to the DFT+U dataset, and with 1:1 energy-to-force weight ratio in the loss function. The cusp regularization factor was not explicitly defined during the training. The simulations for property calculation are performed on a smaller 3 $\times$ 3 $\times$ 3 UO$_2$ structure. Erratic MLIP predictions especially at high temperatures suggests that spectra are needed for both elements U and O.}
    \label{fig:onlyU_re-training_77K_properties}
\end{figure}

\begin{figure}[H]
    \centering
    \includegraphics[width=0.85\linewidth]{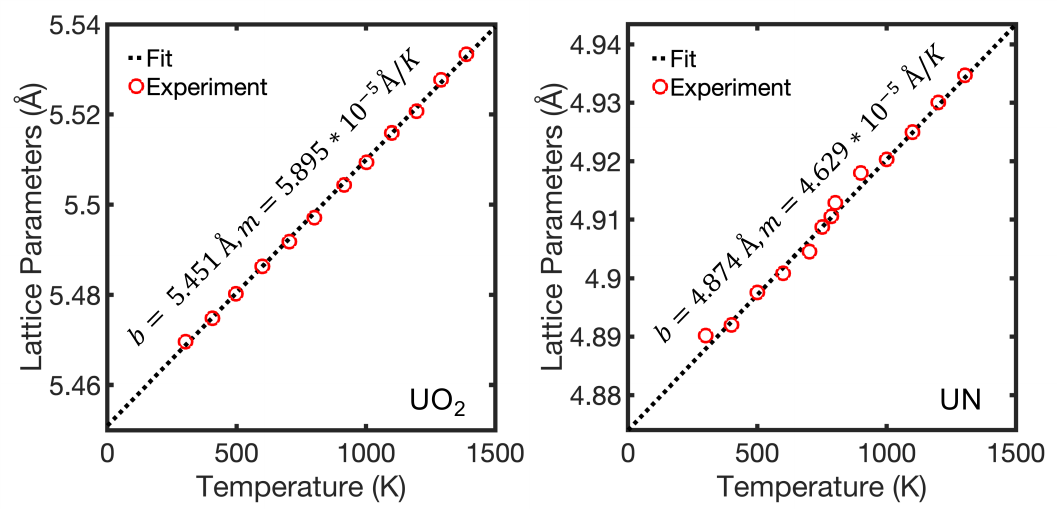}
    \caption{Linear fit ($b+m\times T$) for experimental lattice parameters as a function of temperature to estimate the lattice parameters at the MLIP refinement temperature of UO$_2$ (left) and UN (right). From this fit, the lattice parameter for UO$_2$ at 75 K is 5.4577~\AA, and for UN at 50 K is 4.8763~\AA. This estimation is used to construct the perfect stoichiometric lattice needed to develop synthetic target EXAFS spectra for the refinement process in this work.}
    \label{fig:UO2_UN_lp}
\end{figure}

\begin{figure}[H]
    \centering
    \includegraphics[width=0.7\linewidth]{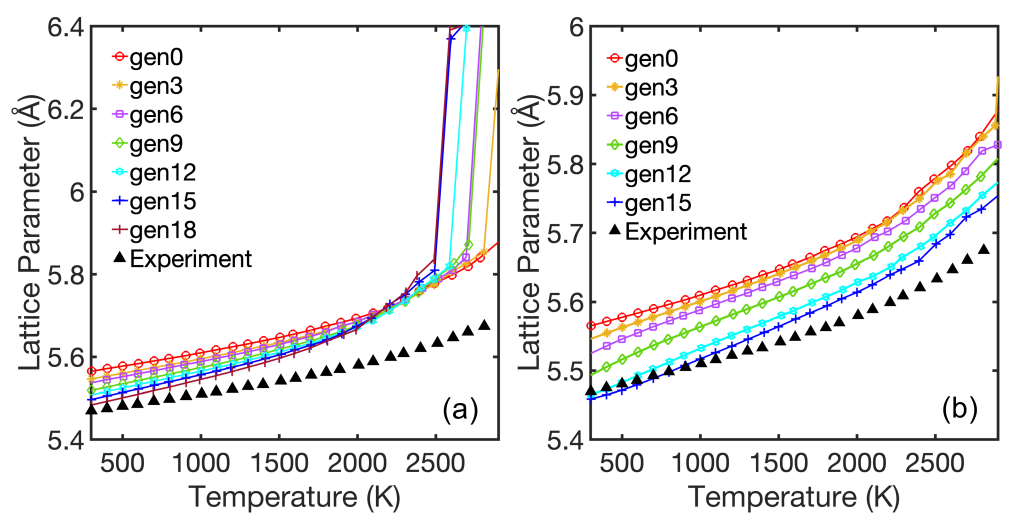}
    \caption{Lattice parameters at re-training iterations when (a) training UO$_2$ MLIP to only the synthetic U spectra, and (b) training UO$_2$ MLIP to only the synthetic O spectra}
    \label{fig:trainingtoSyntheticUandO}
\end{figure}

\begin{figure}[H]
    \centering
    \includegraphics[width=0.7\linewidth]{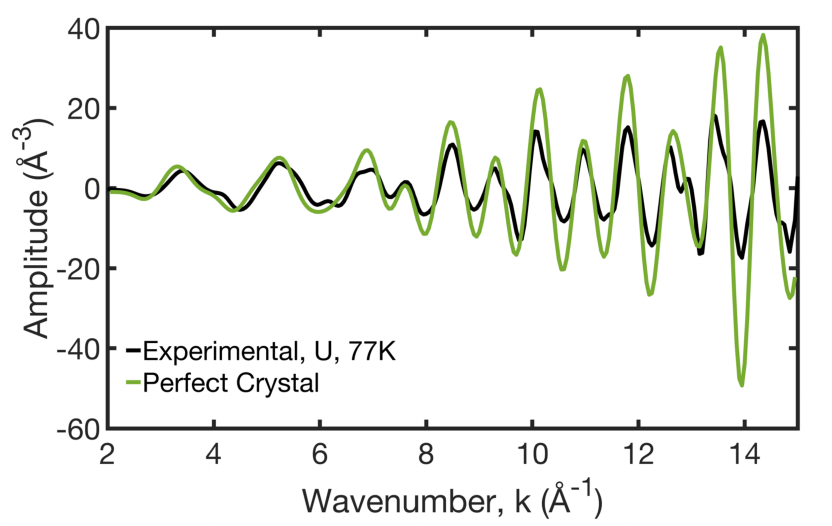}
    \caption{Comparing the experimental EXAFS spectrum for U in UO$_2$ at 77 K with the target synthetic EXAFS spectrum. The target synthetic EXAFS spectrum is computed for a U atom in $4 \times 4 \times 4$ perfect crystal modeled with experimental lattice parameters at 77 K.}
    \label{fig:UO2_exafs_target_n_expt}
\end{figure}

\begin{figure}[H]
    \centering
    \includegraphics[width=0.5\linewidth]{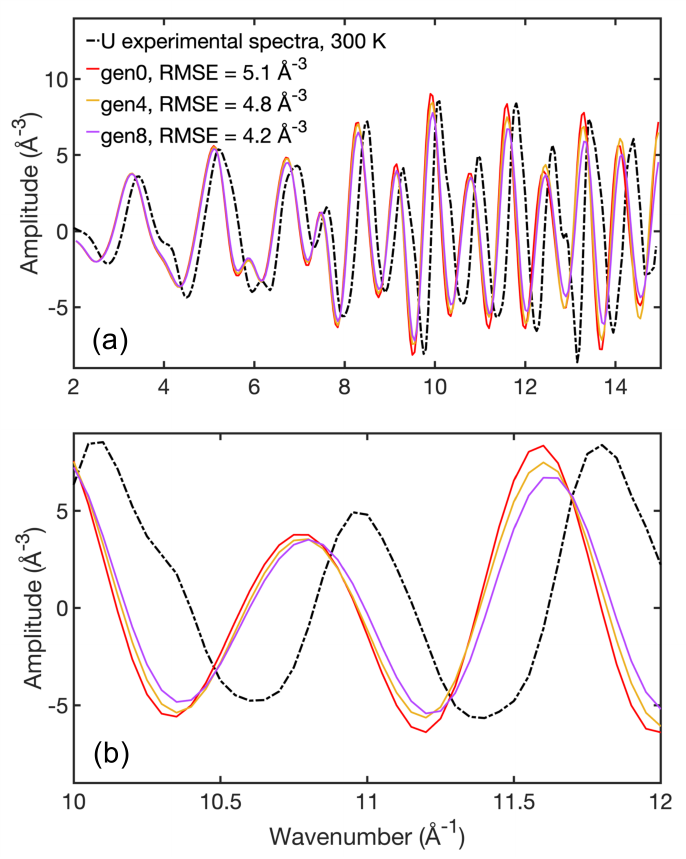}
    \caption{Re-training to match the experimental EXAFS spectra only for U in UO$_2$ at 300~K. In the absence of experimental EXAFS spectra for O, the atomic energy shifts corresponding to oxygen atoms are assumed to be zero. Expensive SCF method in FEFF is used to generate the EXAFS spectra at each generation. Unrefined MLIP had 1:1 energy-to-force weight ratio. Cusp regularization factor for training in HIP-NN was not included.}
    \label{fig:onlyU_re-training_300K}
\end{figure}

\begin{figure}[H]
    \centering
    \includegraphics[width=1.0\linewidth]{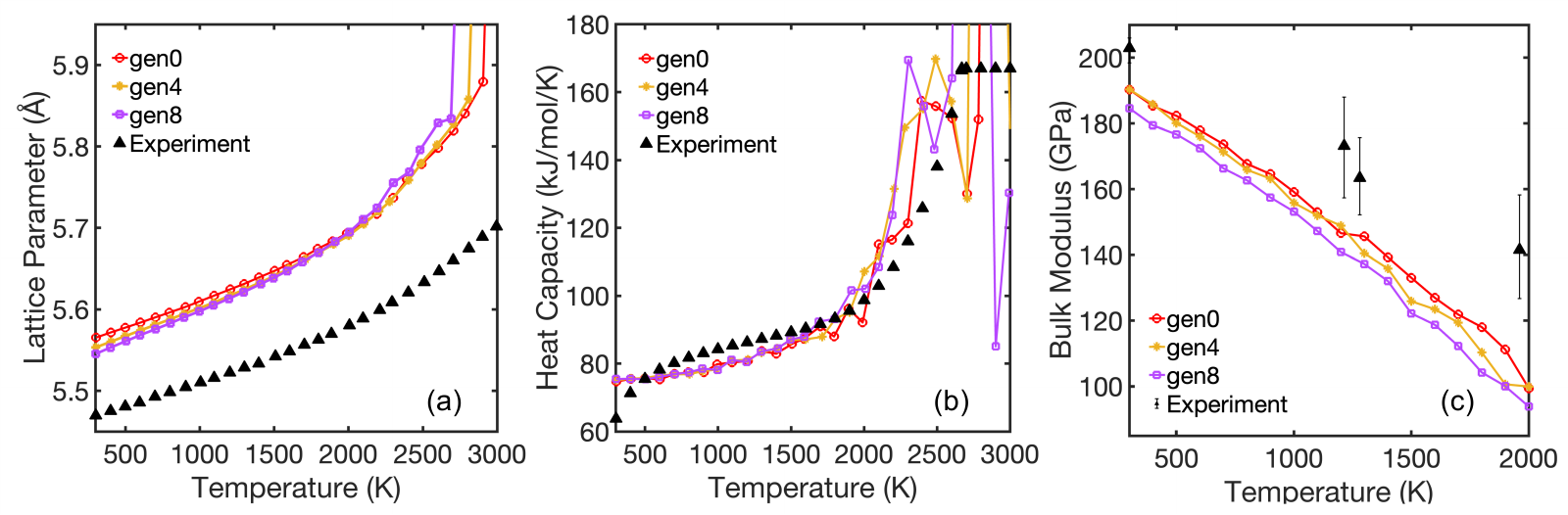}
    \caption{Re-training to match the experimental EXAFS spectra only for U in UO$_2$ at 300~K. The gen0 here corresponds to the unrefined MLIP pre-trained to DFT+U dataset and 1:1 energy-to-force weight ratio, with the default cusp regularization factor. The simulations for property calculation are performed on a smaller 3 $\times$ 3 $\times$ 3 UO$_2$ structure. Erratic behavior is evidence that spectra are needed for both elements U and O.}
    \label{fig:onlyU_re-training_300K_properties}
\end{figure}

\begin{figure}[H]
    \centering
    \includegraphics[width=0.9\linewidth]{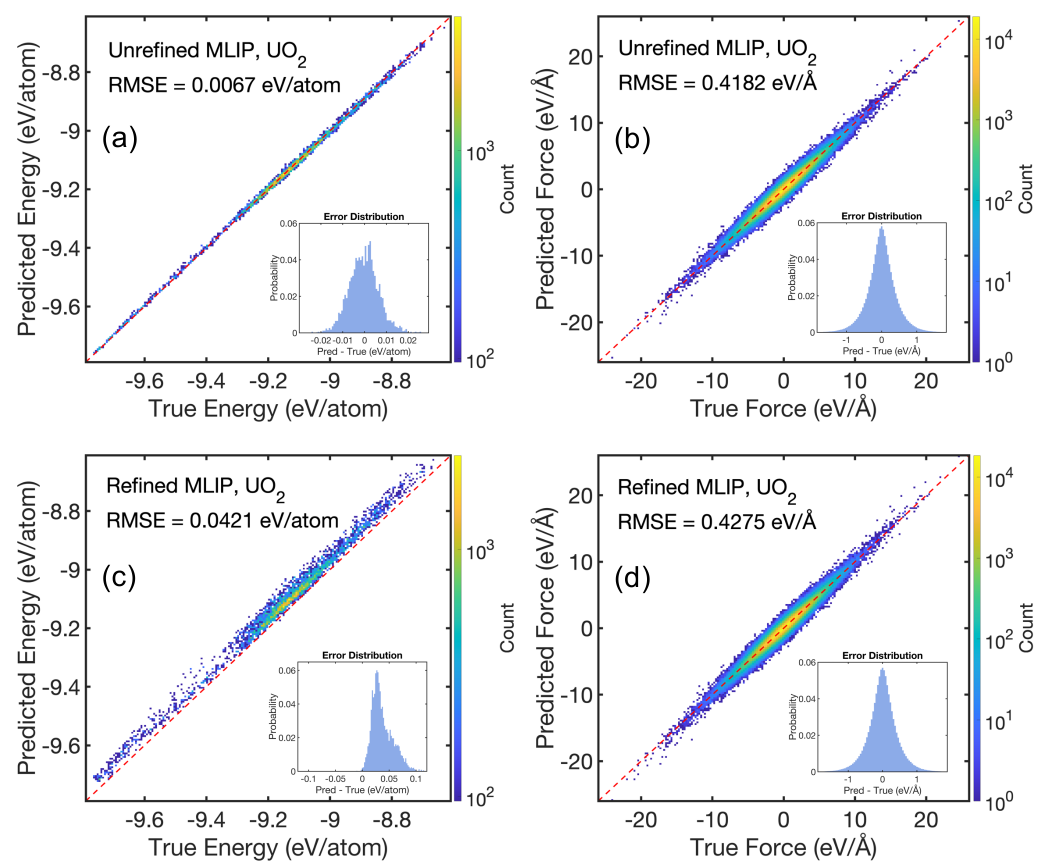}
    \caption{Comparison between MLIP (predicted) energy and forces with DFT+U (true) energy and forces for UO$_2$ for the entire DFT+U dataset. The predicted values in panels (a) and (b) are computed with the unrefined MLIP, while the predicted values in panels (c) and (d) are computed with the refined MLIP. Larger energy errors for refined MLIP are caused by retraining to shifted atomic energies. Nearly all energy errors are positive, demonstrating a systematic shift during refinement. Force errors are similar between the unrefined and refined models, suggesting that the refined MLIP has not forgotten the fundamental physics in the DFT+U dataset.}
    \label{fig:correlationUO2}
\end{figure}

\begin{figure}[H]
    \centering
    \includegraphics[width=0.9\linewidth]{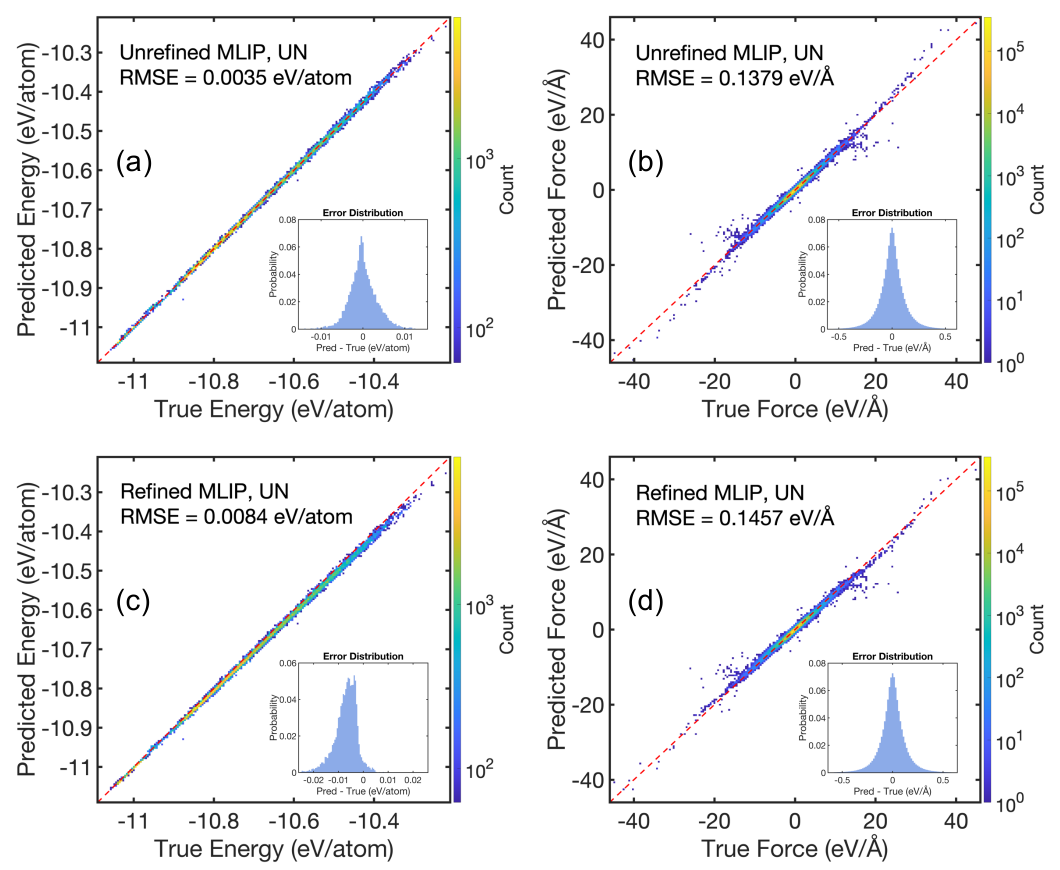}
    \caption{Comparison between MLIP (predicted) energy and forces with DFT (true) energy and forces for UN for the entire DFT dataset. The predicted values in panels (a) and (b) are computed with the unrefined MLIP, while the predicted values in panels (c) and (d) are computed with the refined MLIP. Larger energy errors for refined MLIP are caused by retraining to shifted atomic energies. Nearly all energy errors are negative, demonstrating a systematic shift during refinement. Force errors are similar between the unrefined and refined models, suggesting that the refined MLIP has not forgotten the fundamental physics in the DFT dataset.}
    \label{fig:correlationUN}
\end{figure}

\begin{figure}[H]
    \centering
    \includegraphics[width=0.7\linewidth]{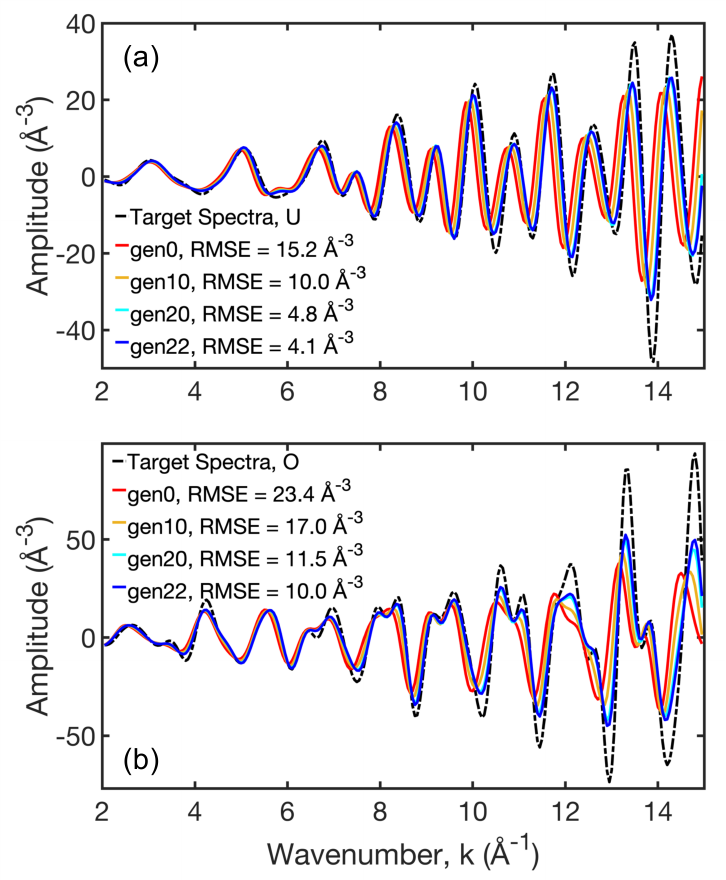}
    \caption{Comparing the EXAFS spectra for (a) U, and (b) O in UO$_2$ computed using unrefined and refined MLIPs with the target EXAFS spectra at 75~K temperature. Here, gen0 corresponds to the unrefined MLIP pre-trained to DFT+U dataset, gen22 corresponds to the final EXAFS-refined MLIP, and, genx with $0<$x$<22$ corresponds to all the other generations during the refinement process. The target EXAFS spectra are synthetic spectra computed for a perfect crystal. Clearly, the agreement of refined MLIP EXAFS with target spectra is better than the unrefined MLIP. Similar trend is observed on comparing the refined MLIP-predicted U EXAFS spectra with the experimental spectra as shown in Figure 2 in main paper.}
    \label{fig:UO2_compare_exafs_with_target}
\end{figure}

\begin{figure}[H]
    \centering
    \includegraphics[width=1.0\linewidth]{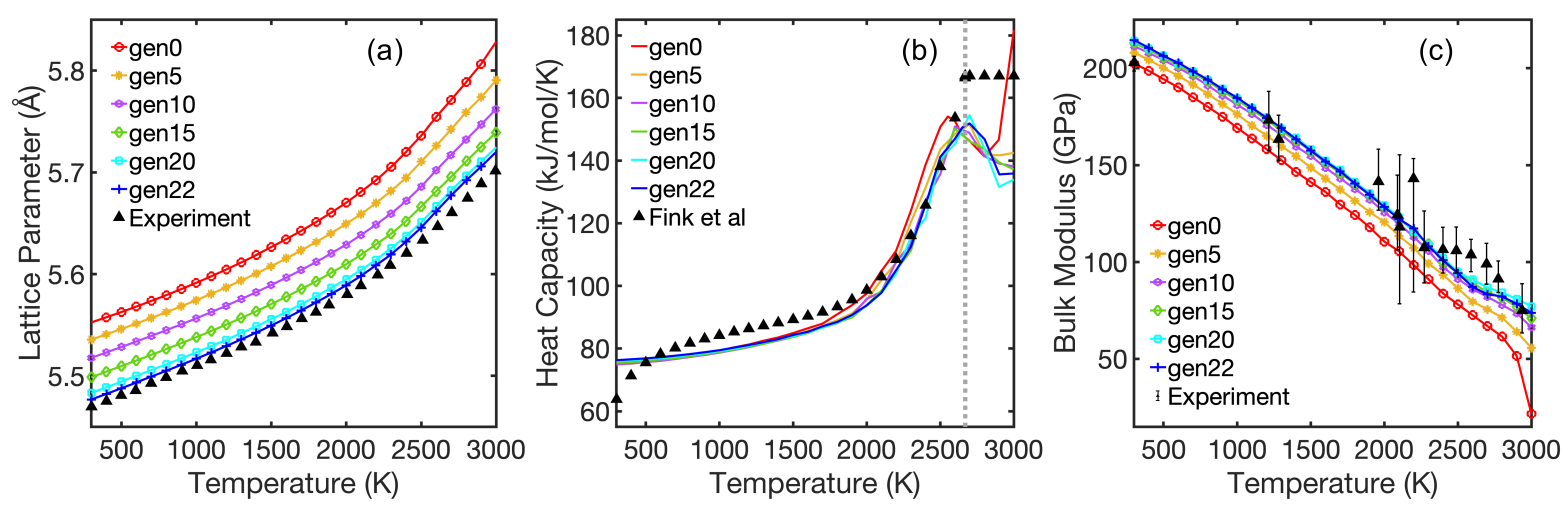}
    \caption{A systematic improvement in the thermophysical properties of UO$_2$ (a) lattice parameters, (b) heat capacity, and (c) bulk modulus at each refinement iteration. Here, gen0 corresponds to the unrefined MLIP pre-trained to DFT+U, gen22 corresponds to the final EXAFS-refined MLIP, and, genx with $0<$x$<22$ corresponds to all the other generations during the refinement process.}
    \label{fig:UO2_properties_atGen}
\end{figure}

\begin{figure}[H]
    \centering
    \includegraphics[width=0.6\linewidth]{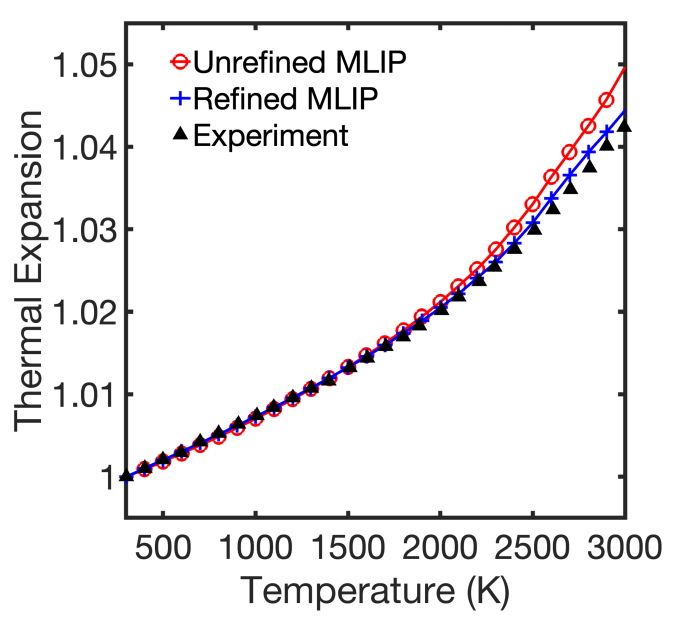}
    \caption{Thermal expansion of UO$_2$ calculated using unrefined and refined MLIPs, compared with the experimental results. Thermal expansion is defined as the lattice parameter at temperature T (a$_{\rm T}$) divided by the lattice parameter at 300~K (a$_{\rm 300K}$).}
    \label{fig:UO2_thermal_expansion}
\end{figure}

\begin{figure}[H]
    \centering
    \includegraphics[width=0.7\linewidth]{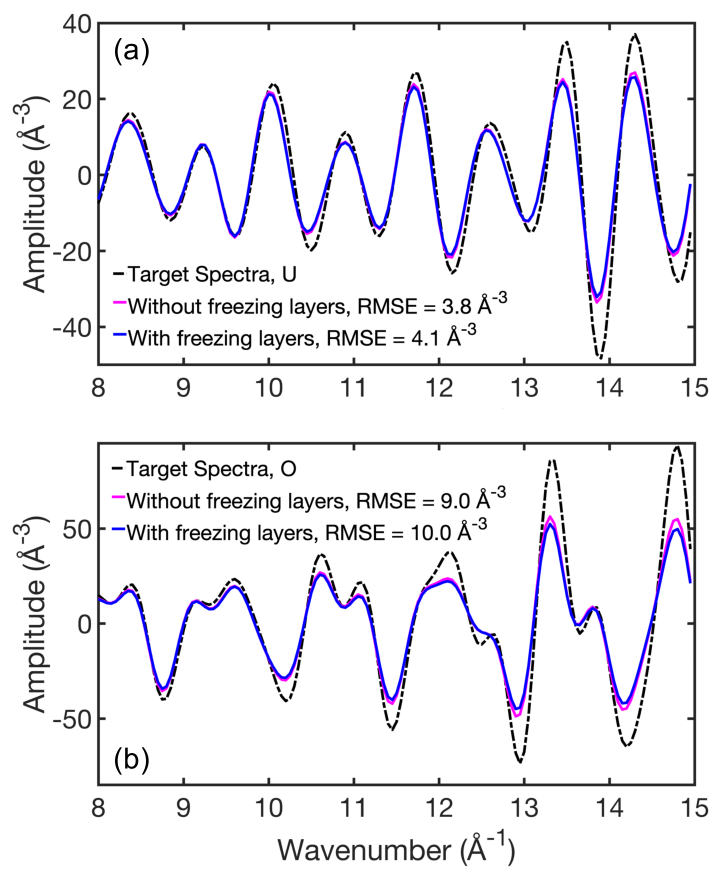}
    \caption{Comparing the target EXAFS spectra with the EXAFS spectra computed using refined MLIPs with and without freezing layers during refinement at 75 K for (a) U and (b) O in UO$_2$. It is observed that the EXAFS spectra have slightly higher amplitude for refined MLIP without freezing layers, compared to the case when layers were frozen during refinement. The spectra are plotted only for wavevector k =8 to 15 to highlight the differences, which are prominent at higher values of k. Notably, only 8 iterations were performed for refining the MLIP when no layers were frozen, compared to 22 iterations when some layers were kept frozen.  While the agreement with target EXAFS spectra is similar for refined MLIPs with and without freezing layers, the corresponding thermophysical properties are significantly different, as shown in Figure~\ref{fig:UO2_noFreezing}.}
    \label{fig:UO2_Freezing_EXAFS_compare}
\end{figure}

\begin{figure}[H]
    \centering
    \includegraphics[width=1.0\linewidth]{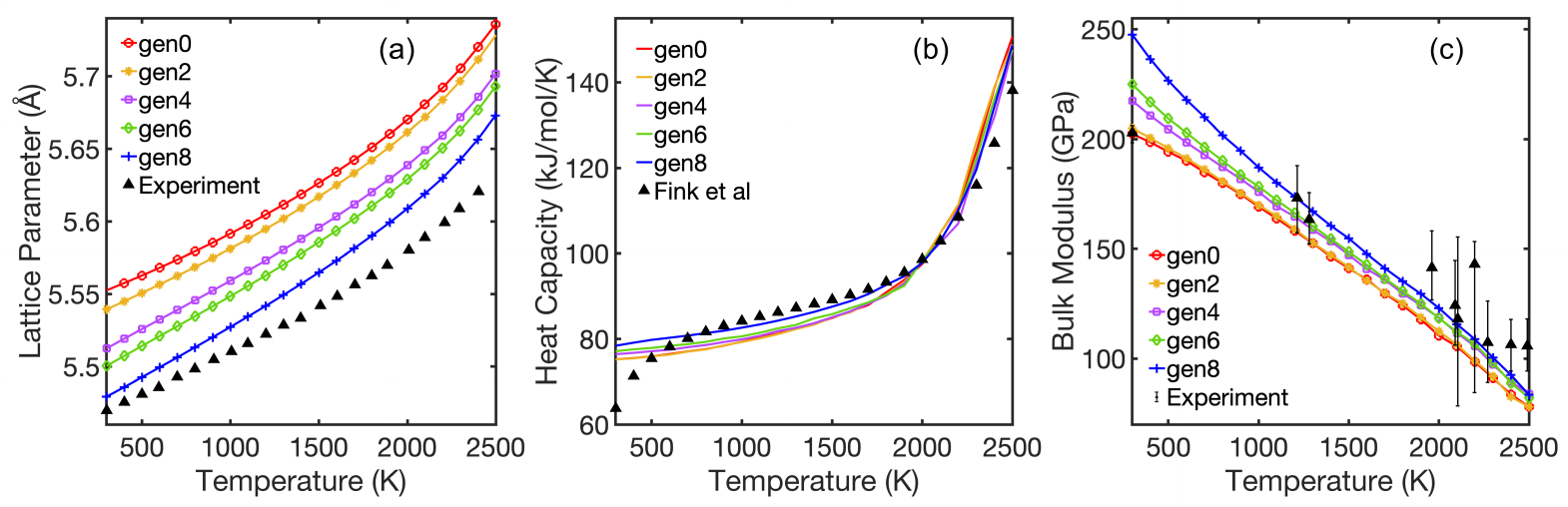}
    \caption{The thermophysical properties of UO$_2$ (a) lattice parameters, (b) heat capacity, and (c) bulk modulus as a function of temperature, for the resultant MLIPs during the refinement process when no layers were frozen in the HIP-NN architecture. The over-predicted bulk modulus at low temperatures and the over-predicted lattice parameters/thermal expansion at high temperatures is evidence of overfitting to the EXAFS spectra. This signifies the importance of freezing the HIP-NN layers when re-training to scarce refinement dataset.}
    \label{fig:UO2_noFreezing}
\end{figure}

\begin{figure}[H]
    \centering
    \includegraphics[width=1.0\linewidth]{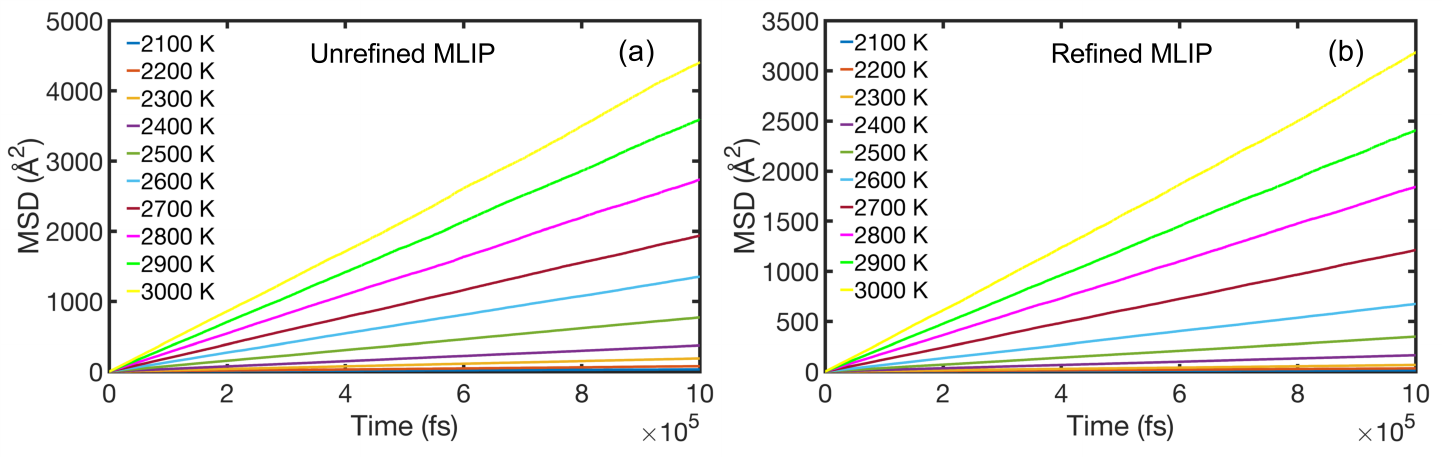}
    \caption{The mean-square-displacement (MSD) of oxygen atoms in UO$_2$ over temperature range of 2100 K to 3000 K with 100 K increments for (a) unrefined MLIP and (b) refined MLIP.}
    \label{fig:UO2_msd}
\end{figure}

\begin{figure}[H]
    \centering
    \includegraphics[width=0.65\linewidth]{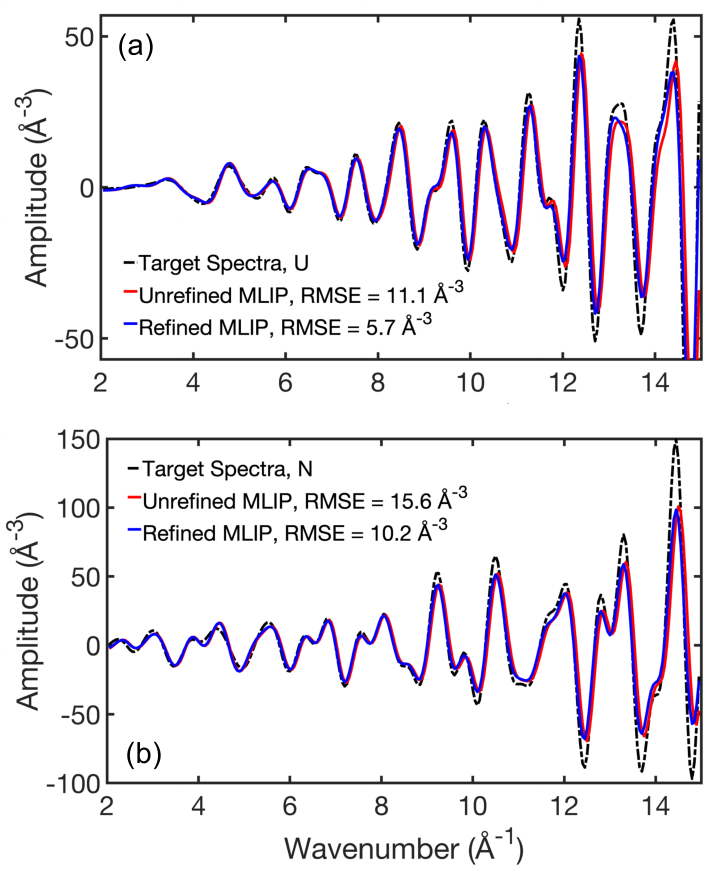}
    \caption{Comparing the EXAFS spectra calculated for unrefined and refined MLIPs with the target EXAFS spectra for (a) U, and (b) N in UN. The unrefined MLIP is trained with energy-to-force pre-factor ratio of 1:1.}
    \label{fig:UN_exafs_comparison_w_Target}
\end{figure}

\begin{figure}[H]
    \centering
    \includegraphics[width=1.0\linewidth]{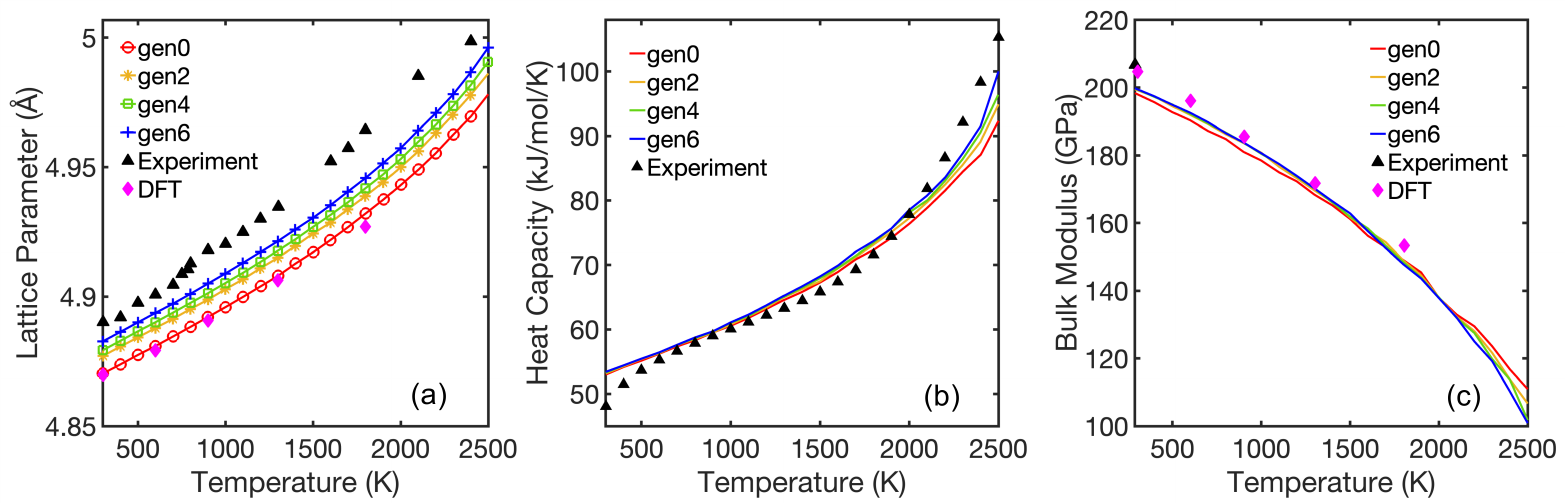}
    \caption{UN thermophysical properties in the case when DFT pre-trained MLIP is trained using a loss function with energy-to-force weights in ratio 1:100 for (a) lattice parameters, (b) heat capacity, and (c) bulk modulus as a function of temperature. A comparison is made for the unrefined MLIP (gen0), with the refined MLIP (gen6), and other generations in between (genx), with the experimental values. To be noted, the bulk modulus and heat capacity show minimal changes during the refinement, compared to the MLIP refinement process for unrefined MLIP trained with 1:1 energy-to-force pre-factor ratio.}
    \label{fig:UN_properties_1-100}
\end{figure}

\begin{table}[H]
\centering
\caption{Point defect energies for UN predicted with unrefined and refined MLIPs in the case when DFT pre-trained MLIP is trained using a loss function with energy-to-force weights in ratio 1:100. Comparison done with the calculated DFT values from literature.}
\label{tab:defectsUN}
\begin{tabular}{@{}lcccccc@{}}
\toprule
Defect energies (eV) & Schottky defect & Frenkel Pair (U) & Frenkel Pair (N) \\ 
\midrule
 DFT  & 5.15 & 9.46 & 5.04 \\
 Unrefined MLIP & 4.81 & 11.22 & 5.82 \\
 Refined MLIP & 5.11 & 10.91 & 5.99 \\
\bottomrule
\end{tabular}
\end{table}

\begin{table}[H]
\centering
\caption{Elastic constants and bulk modulus for UN, calculated with unrefined and refined MLIPs at 290 K, in the case when DFT pre-trained MLIP is trained using a loss function with energy-to-force weights in ratio 1:100. Comparison is done with the experimental values at 290 K.}
\label{tab:elasticsUN}
\begin{tabular}{@{}lcccccc@{}}
\toprule
Elastic constants & C$_{11}$ (GPa) & C$_{12}$ (GPa) & C$_{44}$ (GPa) 
& Bulk modulus (GPa) \\ 
\midrule
 Experiment & 423.9 $\pm$ 0.6 & 98.1 $\pm$ 0.9 & 75.7 $\pm$ 0.2 & 206.7 $\pm$ 0.8\\
 Unrefined MLIP & 380.0 & 109.6 & 47.4 & 199.7 \\
 Refined MLIP & 370.9 & 121.7 & 46.3 & 204.8 \\
\bottomrule
\end{tabular}
\end{table}


\end{document}